\documentclass[sigconf]{acmart}
\usepackage{hyperref}

\AtBeginDocument{%
  \providecommand\BibTeX{{%
    \normalfont B\kern-0.5em{\scshape i\kern-0.25em b}\kern-0.8em\TeX}}}

\setcopyright{acmcopyright}
\setcopyright{rightsretained}
\acmConference[CHI '25]{Proceedings of the 2025 CHI Conference on Human Factors in Computing Systems}{April 26--May 1, 2025}{Yokohama, Japan}
\acmBooktitle{Proceedings of the 2025 CHI Conference on Human Factors in Computing Systems (CHI '25), April 26--May 1, 2025, Yokohama, Japan}
\acmYear{2025}
\acmDOI{10.1145/3613904.3642625}

\usepackage{hyperref}

\usepackage{arydshln}
\usepackage{dirtytalk}
\usepackage{multirow}
\usepackage{xcolor}
\usepackage{cancel}
\usepackage{subcaption}
\usepackage{enumitem}
\usepackage{color}
\usepackage{tabularray}  
\usepackage{graphicx}    
\usepackage{ulem} 
\author{Zhuoyan Li}
\email{li4178@purdue.edu}
\affiliation{
  \institution{Purdue University}
  \city{West Lafayette}
  \state{Indiana}
  \country{USA}
  \postcode{47907}
}

\author{Hangxiao Zhu}
\email{hangxiao@tamu.edu}
\affiliation{
  \institution{Texas A\&M University}
  \city{College Station}
  \state{Texas}
  \country{USA}
  \postcode{06268}
}

\author{Zhuoran Lu}
\email{lu800@purdue.edu}
\affiliation{
  \institution{Purdue University}
  \city{West Lafayette}
  \state{Indiana}
  \country{USA}
  \postcode{47907}
}

\author{Ziang Xiao}
\email{ ziang.xiao@jhu.edu}
\affiliation{
  \institution{Johns Hopkins University}
  \city{Baltimore}
  \state{Maryland}
  \country{USA}
  \postcode{47907}
}

\author{Ming Yin}
\email{mingyin@purdue.edu}
\affiliation{
  \institution{Purdue University}
  \city{West Lafayette}
  \state{Indiana}
  \country{USA}
  \postcode{47907}
}

\begin{document}

\title[]{From Text to Trust: Empowering AI-assisted Decision Making with Adaptive LLM-powered Analysis
%\cl{how about "The Value, Benefits, and Concerns of AI Assistance in Writing: The Role of Generative AI"}
%Is the more the better? Understanding the impact of collective decision in AI-assisted decision-making
}

\renewcommand{\shortauthors}{Li, et al.}
\newcommand{\squishlist}{
   \begin{list}{\small $\bullet$}
    { \setlength{\itemsep}{0pt}      \setlength{\parsep}{1pt}
      \setlength{\topsep}{1pt}       \setlength{\partopsep}{1pt}
     \setlength{\leftmargin}{1.2em} \setlength{\labelwidth}{1em}
      \setlength{\labelsep}{0.5em} } }
\newcommand{\squishend}{  \end{list}  }
\newcommand{\whosays}[3]{\begin{center}\textit{\say{#3}} (Subject\##1, Group\##2)\end{center}}
\begin{abstract}
AI-assisted decision making becomes increasingly prevalent, yet individuals often fail to utilize AI-based decision aids appropriately especially when the AI explanations are absent, potentially as they do not %understand 
reflect on AI's decision recommendations critically. Large language models (LLMs), with their exceptional conversational and analytical capabilities, present great opportunities to enhance AI-assisted decision making in the absence of AI explanations by providing natural-language-based analysis of AI's decision recommendation, e.g., how each feature of a decision making task might contribute to the AI recommendation.  In this paper, via a randomized experiment, we first show that presenting LLM-powered analysis of each task feature, either sequentially or concurrently, does not significantly improve people's AI-assisted decision performance. To enable decision makers to better leverage LLM-powered analysis, we then propose an algorithmic framework to characterize the effects of LLM-powered analysis on human decisions and dynamically decide which analysis to present. Our evaluation with human subjects shows that this approach effectively improves decision makers' appropriate reliance on AI in AI-assisted decision making.
\end{abstract}

\begin{CCSXML}
<ccs2012>
   <concept>
       <concept_id>10003120.10003121.10011748</concept_id>
       <concept_desc>Human-centered computing~Empirical studies in HCI</concept_desc>
       <concept_significance>500</concept_significance>
       </concept>
   <concept>
       <concept_id>10010147.10010178</concept_id>
       <concept_desc>Computing methodologies~Artificial intelligence</concept_desc>
       <concept_significance>500</concept_significance>
       </concept>
 </ccs2012>
\end{CCSXML}

\ccsdesc[500]{Human-centered computing~Empirical studies in HCI}
\ccsdesc[500]{Computing methodologies~Artificial intelligence}

\keywords{AI-assisted decision making, Explainable AI, Large language model}
% \keywords{Human-AI interaction, Group-AI interaction, Diversity}


\maketitle
\section{Introduction}

AI systems have been significantly integrated into the human decision making process in various domains, such as criminal justice~\cite{ghasemi2021application,wang2023pursuit} and financial investment~\cite{khan2022stock,ashtiani2023news}, thereby creating a new paradigm of human-AI collaboration~\cite{wang2020human}. In this paradigm, AI models provide recommendations or analysis to assist humans in making decisions, while human decision makers are ultimately responsible for the final decisions~\cite{jarrahi2018artificial,reverberi2022experimental}.

\begin{table*}[t]
\centering
\small
\begin{tblr}{
  width = \textwidth,
  colspec = {Q[1]Q[3]},
  row{odd} = {c},
  cell{1}{1} = {c=2}{\textwidth},
  cell{3}{1} = {c=2}{\textwidth},
  hline{1,2,3,4,5} = {-}{},
  vline{2} = {2,4}{},
}
\textbf{Income Prediction} &  \\
\textbf{Education Level} & {\smaller 
  \uline{\textit{Value:}}~10 years\\
  \uline{\textit{Concept:}}~Higher education is commonly linked to higher earning potential.\\
  \uline{\textit{In this case:}}~With 10 years of education, this might be slightly below the threshold for high-earning positions, which~\textbf{\textcolor{red}{decreases}}~the likelihood of making over \$50000 per year.
} \\
\textbf{Recidivism Prediction} &  \\
\textbf{Charge Degree} & {\smaller 
  \uline{\textit{Value:}}~Felony\\
  \uline{\textit{Concept:}}~The severity of the charge can predict recidivism, with felonies often leading to harsher predictions than misdemeanors.\\
  \uline{\textit{In this case:}}~Facing a felony charge, which~\textbf{\textcolor[rgb]{0,0.502,0}{increases}}~the likelihood of recidivating because felonies are associated with more severe criminal behavior.
} 
\end{tblr}

\caption{Examples of analysis generated by the LLM for (top) how the education level of a person might have affected AI's prediction on this person's income level; and (bottom) how the charge degree of a defendant might have affected AI's prediction on this defendant's recidivism status.}
\label{tab:llm_rationales}
\vspace{-15pt}
\end{table*}

However, many empirical studies evaluating the effectiveness of current AI-assisted decision making systems ~\cite{lai2021towards} have demonstrated that when people collaborate with AI in decision making tasks, they rarely engage in analytical thinking to combine their own insights with the AI model's recommendations intelligently~\cite{he2023knowing,prabhudesai2023understanding,buccinca2021trust}. Instead, they %frequently inappropriately 
often rely on AI inappropriately---accepting an AI model's recommendations when they are incorrect or mistakenly ignoring AI's correct recommendations---leading to either overreliance or underreliance on AI~\cite{schemmer2022should,lu2021human}. 
To address this problem, 
previous research~\cite{schemmer2023appropriate,schoeffer2024explanations,alufaisan2021does} proposed to display explanations generated by post-hoc explainable AI (XAI) methods~\cite{ribeiro2016should,lundberg2017unified, huang2023can} along with the AI model's decision recommendations to assist people in evaluating AI's reliability and identifying optimal strategies for relying on AI. However, the computation of AI explanation often requires access to the AI model's internal parameters and structures, while many evaluation studies have revealed that it is challenging for humans to understand and utilize such explanation without substantial effort to teach them how to interpret the explanation~\cite{wang2021explanations}. Consequently, 
even with the presence of AI explanations, decision makers often still exhibit a low level of appropriate reliance on AI, let alone the case when the AI explanations are not available.

As such, one would naturally wonder if it is still feasible to guide decision makers to critically and systematically reflect on AI's decision recommendations and appropriately utilize them \textit{without easily accessible or available AI explanations}. Interestingly, in real-world decision making across various domains like healthcare and finance, when decision makers find the initial recommendations lack transparency and clarity, they often seek additional insights or interpretations from secondary sources. To this end, the exceptional conversational and analytical capabilities exhibited by the latest state-of-the-art large language models (LLMs) could offer strong promise~\cite{zhao2023survey,clark2021all,li2024pre,10.1145/3613904.3642482,10.1145/3613904.3642400,karinshak2023working}. For example, LLMs can analyze a decision making task and AI's decision recommendation on it, and then provide potential reasons for why the AI recommends such a decision in a natural language format, which is straightforward for humans to process.
%(such as those shown in Table~\ref{tab:llm_rationales}). 
%Such natural-language-based analyses can be easier for people to interpret than those highly technical AI explanations generated by XAI methods such as feature importance plots.  
While the AI model serves as the primary advisor for human decision makers, when LLM-powered analyses are used to augment the AI model's recommendations, the LLM effectively serves as a secondary advisor to provide additional perspectives and justifications through its analysis. 
These analyses may help the human decision maker better interpret the recommendation of the primary advisor.  %and explore different viewpoints. 
They may also offer a starting point for decision makers to organize their thoughts and reflect on both the AI model's decision recommendation and their own judgment, which may help them calibrate their trust in the AI model.

Therefore, in this paper, we start by conducting a randomized human-subject experiment to examine whether incorporating LLM-powered analyses in AI-assisted decision making can improve the performance of human-AI teams and promote more appropriate reliance on AI models in the absence of actual explanations of the AI models. %Based on OpenAI's GPT-4 model~\cite{openai2023gpt4}, 
Given a decision making task as well as an AI model's decision recommendation on it, we first prompted OpenAI's GPT-4 model~\cite{openai2023gpt4} to %internalize the AI recommendation as its own decision and %analyze the decision making task and 
generate analyses for how each feature of the task might have 
led to the AI model's decision recommendation on the task %influenced the 
%decision recommendation given by the AI model on this task by ``internalizing'' the AI model recommendation as its decision 
(see Table~\ref{tab:llm_rationales} for examples). Depending on whether and how to present these LLM-powered analyses, we created three treatments in our experiment---\textsc{Control} (where participants would not receive any analysis from the LLM), \textsc{Seq} (where participants receive the analysis about each feature sequentially), and \textsc{All} (where participants receive analyses about all features at once). Our experimental results show that presenting LLM-powered analysis either sequentially or concurrently to human decision makers does not significantly improve their decision accuracy compared to those decision makers who did not receive any LLM-powered analysis. This suggests 
that more intelligent ways should be used to present LLM-powered analyses to people to facilitate their utilization of this information and promote their effective decision making.

In light of this, we propose an algorithmic framework to adaptively present LLM-powered analyses to decision makers---based on the historical data on how human decision makers react to different LLM-powered analyses, our algorithmic framework learns to present LLM-powered analysis selectively and progressively to maximize the chance for the decision maker to rely on the AI model's decision recommendations appropriately and make the correct decisions. 
To do so, we first learn a human behavior model that characterizes the effects of LLM-powered analysis on human decisions. We then dynamically decide which analysis to present (among the LLM-powered analyses for all features of the decision making task) by comparing the expected maximum utility of presenting each analysis. 
To evaluate the effectiveness of this algorithmic approach in selecting the best set of LLM-powered analyses to help improve decision makers' appropriate reliance and decision accuracy in AI-assisted decision making, we conducted another randomized human-subject experiment. 
We find that compared to other baseline approaches for presenting LLM-powered analysis, when the LLM-powered analyses are selected using our algorithmic approach, human decision makers can achieve significantly higher accuracy in their final decisions and reduce overreliance on the AI model across different types of decision making tasks. Additional analysis suggests that our algorithmic approach selects fewer but more informative LLM-powered analysis to show to decision makers compared to baseline approaches. 

Together, our study provides important experimental evidence regarding the effectiveness of incorporating LLMs in AI-assisted decision making, and how to design intelligent interactions between humans and LLMs to %fully leverage the potential of LLMs for 
promote better human-AI collaboration in decision making. 
We conclude by discussing the implications and limitations of our study.

\section{Related Work}
\label{sec:related}
%\my{Beef up this section...could cover some grounds about AI explanations, and previous studies that highlight explanations need to be provided selectively/progressively, etc.}

\subsection{AI-Assisted Decision Making}
The increasing prevalence of AI-assisted decision making has led to a growing line of research to investigate how people engage with, trust in, and rely on AI models in this new collaboration paradigm ~\cite{lai2021towards,chiang2023two,10.1145/3653708}. Early studies focus on empirically identifying factors that influence AI-assisted decision making, including the AI model's performance~\cite{rechkemmer2022confidence}, the explanation of the model recommendation~\cite{schemmer2023appropriate,schoeffer2024explanations,robbemond2022understanding},  the decision making workflow~\cite{chiang2024enhancing,rastogi2022deciding}, and the influence of task complexity on human-AI interactions~\cite{salimzadeh2023missing}.

While it is expected that the complementarity between AI models and humans could enable the human-AI team to outperform either party alone, in practice, the collaboration between humans and AI in decision making is widely observed to be suboptimal ~\cite{schemmer2022should}. It is observed that people usually exhibit inappropriate reliance on AI models~\cite{steyvers2023three}. For instance, the design of conversational interfaces can influence users' trust, sometimes causing overreliance on AI recommendations~\cite{gupta2022trust}. In addition, people may also blindly rely on AI in time-pressured environments, where the presence of AI suggestions may speed up decision making at the cost of accuracy~\cite{swaroop2024accuracy}. In contrast, people could also reject the AI model recommendation even when it is correct, noted as underreliance on AI~\cite{mahmud2022influences,ochmann2020influence,10.1145/3648188.3675130}. Recent research has also discussed how misaligned AI outputs can contribute to people's underreliance on AI systems despite their accuracy~\cite{guerdan2022under}. To help decision makers interact with and rely on the AI model more appropriately, a wide range of approaches was recently developed ~\cite{he2022walking,10.1145/3610219,10.1145/3555572,li2023modeling,lu2023strategic,li2024decoding,lu2024mix}. For instance, the cognitive forcing function encourages people to engage with AI more cognitively, thus potentially reducing people's overreliance on the AI model~\cite{buccinca2021trust,erlei2020impact,lai2021towards,salimzadeh2024dealing}. ~\citeauthor{ma2023should}~\cite{ma2023should} explored the calibration of user trust in AI-assisted decision making by inferring the correctness likelihood of both human and AI on a decision case, which informs the adaptive presentations of the AI model's decision recommendations. 
% \my{Discuss Shuai Ma's CHI 2023 paper here.}

In addition, providing AI explanations generated by various post-hoc explainable AI (XAI) methods~\cite{10.1145/2939672.2939778,10.5555/3295222.3295230} that reveal the decision rationale of AI models is another popular approach used, %with the 
aiming to improve humans' understanding of the AI model's behavior and enable humans to calibrate their trust in AI. 
%, which in turn allows users to calibrate their trust in the AI model accordingly. 
However, many empirical studies have observed that people often struggle to process and comprehend these explanations~\cite{vasconcelos2023explanations,lee2023understanding,wang2021explanations,li2024utilizing}, letting alone utilize the insights revealed from these explanations to trust AI more appropriately. %which fails to meet designers' expectations of positively influencing human engagement with AI models. 
To realize the positive utility of explanations in AI-assisted decision making, recent research highlights the need to provide explanations selectively or progressively to aid human comprehension~\cite{feng-boyd-graber-2022-learning,10.1145/3610206,springer2020progressive,springer2019progressive,li2024utilizing} . For instance, ~\citeauthor{10.1145/3610206}~\cite{10.1145/3610206} demonstrated that selectively highlighting AI explanations, which align with the user’s own decision rationale, can increase agreement between human decisions and AI model predictions and reduce human overreliance on AI recommendations. ~\citeauthor{springer2019progressive}~\cite{springer2019progressive} showed that users may benefit from
initially simplified feedback that hides potential AI system errors and assists users in building working heuristics about how the AI system operates progressively. In this work, we make an initial attempt to explore that \textit{in the absence of AI explanations}, whether the incorporation of the natural-language-based, LLM-powered analysis of the AI recommendations on decision making tasks %as a  explanation 
can promote more appropriate reliance behavior of humans on AI models in decision making, and how to present such analysis in the most effective way.

% explanations are a popularly employed method to improve joint decision making via revealing the AI model decision rationales to humans~\cite{vasconcelos2023explanations,lee2023understanding}\hx{, and research has shown that task complexity plays a key role in shaping trust and reliance on AI systems, requiring tailored interventions to manage uncertainty and reliance behavior~\cite{salimzadeh2024dealing}}. Moreover, a series of interventions~\cite{erlei2020impact,lai2021towards} are designed to influence humans in AI-assisted decision making\cite{lai2021towards}.  
%Well-designed training process, as another approach, also hold its promise to guide human decision makers better utilize the AI model~\cite{erlei2020impact}. Despite the great insight provided, the effectiveness of the approaches are either mixed (e.g., in price of under-reliance) or could only be applied to a specific population (e.g., AI literacy). Thus, there has been an universal approach for appropriate and effective AI-assisted decision making.

\subsection{Human-LLM Interaction}

% \my{You should discuss the TalkToModel paper somewhere in this section.}
Recently, large language models (LLMs) have demonstrated their exceptional capabilities across various applications to assist humans, including creative writing~\cite{wang2024weaver,yuan2022wordcraft,li2024value}, software engineering~\cite{nam2024using,ni2023lever}, and generative design ~\cite{huang2024graphimind}, which has sparked significant interest within the HCI community to investigate the interaction between humans and LLMs ~\cite{gao2024taxonomy,kim2024understanding,10.1145/3613904.3642002,chiang2024enhancing,li-etal-2024-disclosure}. On the one hand, LLMs are increasingly utilized to directly create content or solve problems, which is shown to match or even surpass humans' performance. For example, \citeauthor{10.1145/3544548.3581225}~\shortcite{10.1145/3544548.3581225} presented the framework leveraging LLMs to create coherent scripts and screenplays with humans in the loop. 
% \my{What do you mean by ``with humans audition''?}. 
In other cases, LLM-based services provide foundational support for human creation, such as generating coding schemes for qualitative analysis~\cite{chew2023llm}. 
In these human-LLM collaboration scenarios, a key challenge is that laypeople often lack the skill to effectively prompt LLMs to generate the outputs that they desire ~\cite{zamfirescu2023johnny}. To address this challenge, novel approaches like AI Chains~\cite{wu2022ai}, automatic prompting methods ~\cite{shin2020autoprompt}, and interactive interfaces~\cite{wang2024lave,liu2024make} are developed to enhance the effectiveness of human-LLM interaction, either by improving LLMs' usability ~\cite{hong2024next,yang2024human} or by guiding humans' engagement with LLMs.

%\my{To update.}
%However, effective human-LLM interaction requires careful design to optimize usability and outcomes. Without this, such collaborations can lead to suboptimal results. 
%Exemplified by the overreliance on LLM led 
Researchers have also explored the potential of LLMs in AI-assisted decision making. For example, LLMs could directly provide decision recommendations. However, it was found that the 
overconfident and seemingly convincing LLM outputs can mislead people to believe them to be correct~\cite{steyvers2025large} and result in people's overreliance on LLM~\cite{do2024facilitating,kim2024m}. 
%, as the convincing yet incorrect outputs can potentially mislead users who may rely on them as if they were correct~\cite{kim2024m}. 
%\my{I'm not sure what does this sentence mean.}. 
%\my{This sentence is not grammatically correct.}. 
Recently, \citeauthor{slack2023explaining}~\cite{slack2023explaining} developed an interactive dialogue system that allows users to inquire about the reasons behind the AI model's predictions. This system leverages a LLM to parse user intent and match it with pre-specified, handcrafted answers, demonstrating significant potential to enhance user understanding and decision performance through conversational interactions with the AI model.
% \my{So in their paper, only human understanding is improved, not human decision performance? How is our work different from theirs? (Or what makes our setting different so that in the first experiment, we find providing LLM-powered analysis is not that helpful?)} \zy{they also see the improvements over the decision accuracy... In their work, the users have more freedom to chat with the system like why this prediction happens, how one feature affects the prediction, and what the typical errors the AI model would have; in addition, they used LLM mainly to identify user intent, they do not have LLM directly analyze the task and provide analysis. Their language-format explanations are directly parsing from the post0-hoc explanation. }. \my{Then, try to include these points in the text (in a succinct manner.} 
Different from the previous work, %which mostly relies on the heuristic design of human-LLM interaction flow, 
in this paper, we explore 
how to utilize LLMs to analyze an AI model's decision recommendations and augment them, and 
how to build an algorithmic framework to dynamically decide what information to present to humans from the rich information generated by LLMs.
%, thereby yielding enhanced human-AI team performance in decision making.

\section{Empirical Examinations of the Impacts of LLM-Powered Analysis in AI-assisted Decision Making}
\label{empirical_work}
We start by investigating whether the incorporation of LLM-powered analysis can enhance human decision makers' decision performance and promote their more appropriate reliance on AI models in AI-assisted decision making. To do so, we conducted a randomized human-subject experiment on Prolific.

\subsection{Decision Making Task}
In our experiment, we considered two types of decision making tasks that have widely been used as the testbeds in AI-assisted decision making research~\cite{ma2023should,zhang2020effect,chiang2024enhancing}:
\begin{itemize}
    \item \textbf{Income Prediction}~\cite{misc_census_income_20}: Human decision makers were asked to determine whether a person's annual income level is higher or lower than \$50k with the assistance of an AI model. Specifically, in each task, we presented the participant with a person's profile containing 7 features, which include the person's gender, age, education level, marital status, occupation, work type, and working hours per week. We trained a random forest model to provide decision recommendations, and the accuracy of the model was $76\%$.

    \item \textbf{Recidivism Prediction}~\cite{dressel2018accuracy}: Human decision makers needed to predict whether a defendant would reoffend within two years. Each task presented a defendant profile with 8 features, including basic demographics (e.g., gender, age, race), criminal history (e.g., the count of prior non-juvenile crimes, juvenile misdemeanor crimes, and juvenile felony crimes committed), and information related to their current charge (e.g., charge issue, charge degree). We again trained a random forest model to provide decision recommendations, and the accuracy of the model was $62\%$\footnote{For the random forest models used in both the income prediction and recidivism prediction tasks, we used grid search to fine-tune the model parameters such as the depth of the tree and the number of trees. The relatively low level of prediction accuracy of the random forest model was primarily due to the inherent difficulty and uncertainty of the task. We also experimented with using zero-shot and few-shot approaches to  prompt the state-of-the-art LLM, GPT-4, to directly provide binary recommendations on these tasks. When evaluating on the same test dataset, we found that the accuracy of GPT-4 on income prediction tasks and recidivism prediction tasks were 59\% and 56\%, respectively, which were lower than the random forest models.}. 
    %and used grid search to fine-tune model parameters like tree depth and the number of trees.

\end{itemize}

\subsection{Generation of LLM-powered Analysis}

We used LLMs to generate an analysis for each AI-assisted decision making task. Specifically, we prompted GPT-4 to analyze the decision making task and the AI model's decision recommendation. The prompts for GPT-4 to generate the analysis for both the income prediction and recidivism prediction tasks consist of three parts:
% \zx{move to appendix or supplementary materials? } \zy{Just to take up some space here, as the length of our paper is not as long as those typical CHI paper (:}
\begin{itemize}
    \item \textbf{Introduction Prompt}: \texttt{
Please take on the role of a data analyst and prepare to analyze the provided task instances. Your task is to explain how the features in the presented task instances contribute to the AI model predictions provided. Each profile includes various features that you will need to consider in your analysis, [INTRODUCE THE FEATURE NAMES AND DESCRIPTIONS].}
\item   \textbf{Instruction Prompt}: \texttt{
For each presented task, assess how each feature might contribute to [AI MODEL PREDICTION]. For each task, your analysis should contain 1 identifier (index), [NUMBER OF FEATURES] concepts (explanations of how the features could support the model prediction), and [NUMBER OF FEATURES] case descriptions (specific explanations of how the feature values in the current profile support the model prediction). [AN EXAMPLE OF THE OUTPUT ANALYSIS].}
\item  \textbf{Emotional Stimuli Prompt}: 
\texttt{This is an academic study aimed at enhancing human trust in AI system advice through reasonable explanations. The knowledge gained will help improve human-AI collaboration. This mission is critical to the whole human society. Please analyze the task instance thoroughly and provide diverse insights.}
\end{itemize}
The LLM examined the task features and determined how each feature may have contributed to the AI model's prediction. Consequently, the LLM generated a set of analyses for each task instance, associating each task feature with one analysis to indicate its possible contribution to the AI recommendation. Table~\ref{tab:llm_rationales} shows several examples of analyses generated by GPT-4.
This set of analyses serves as the LLM-powered analysis to be incorporated into AI-assisted decision making in our study (see the supplemental materials for more examples of the analyses). 
While such LLM-powered analysis can be readily applied to decision making tasks with tabular data where the task-related information is presented in a structured manner as a collection of features and their values, similar analysis can also be conducted on decision tasks involving other types of data (e.g., images, texts) after transforming the unstructured data into structured formats (see more discussions on this in Section~\ref{sec:generalization}).

Note that we do not consider the analysis generated by the LLM as necessarily reflecting the AI model's true decision rationale. Instead, it is only the LLM's \textit{interpretation/justification}  of the AI recommendation, and is used to augment the AI recommendation in the absence of explanations to the AI model. Alternatively, since we prompted the LLM to justify a specific decision (i.e., the decision that is consistent with the AI model's recommendation), one can also view the LLM-powered analysis as the LLM's own explanations to the specified decision.

\begin{figure*}[t]
  \centering
  \subfloat[\textsc{Seq} treatment]{\includegraphics[width=0.49\textwidth]{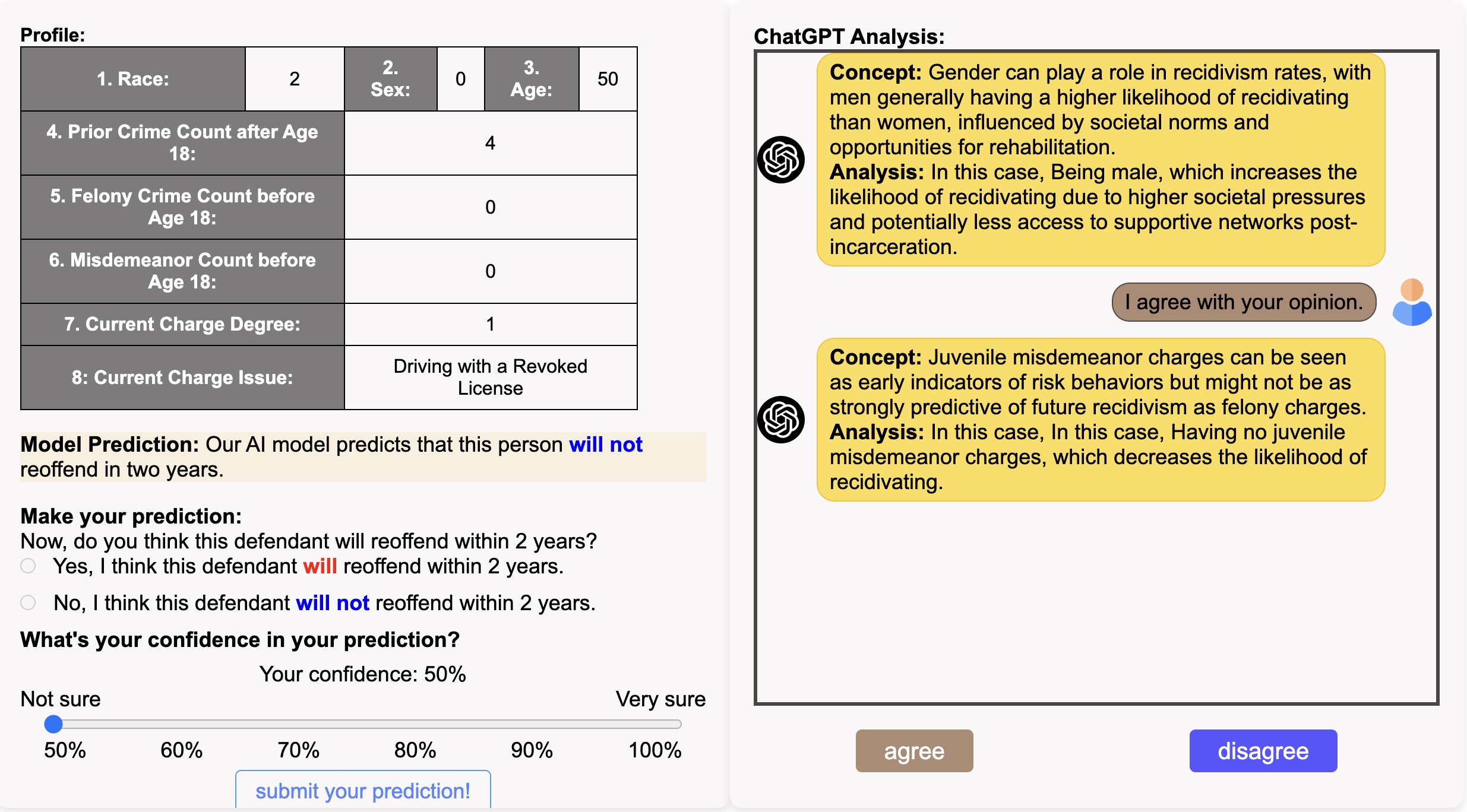}\label{interface_seq}}
  \hfill
  \subfloat[\textsc{All} treatment]{\includegraphics[width=0.49\textwidth]{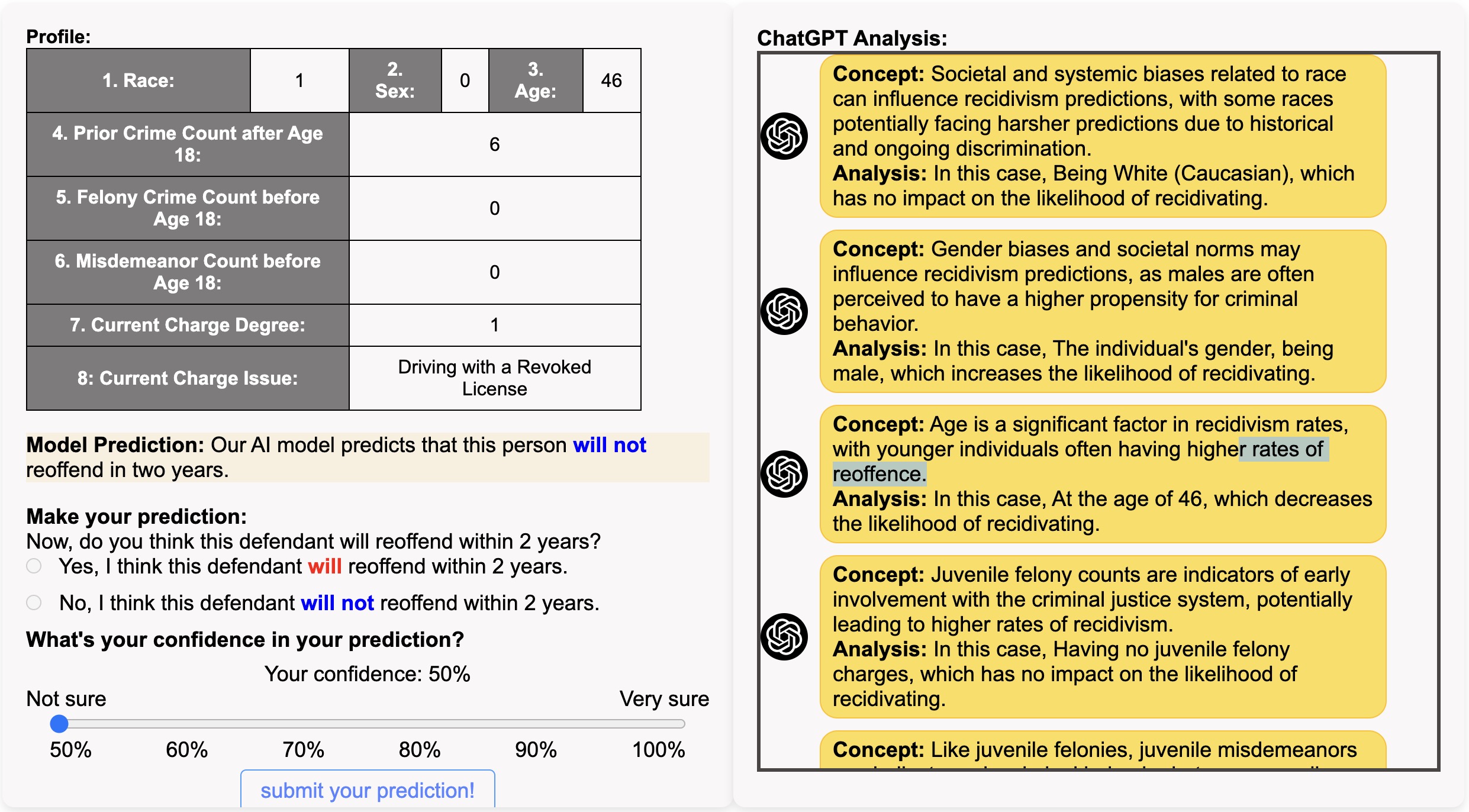}\label{interface_all}}
\caption{The example interfaces used in the \textsc{Seq} and \textsc{All} treatments of our experiment for the recidivism prediction task.
}
\label{fig:interface}
% \vspace{-10pt}
\end{figure*}

\subsection{Experimental Treatments}
\label{exp_treatment}
In our experiment, participants were asked to complete a series of decision making tasks. For each task, they were provided with the AI model's prediction along with the task instance, and they needed to make the final decision.
We created 3 experimental treatments by varying whether and how LLM-powered analysis was introduced into the AI-assisted decision making process.  Specifically: 
%In addition, we created the following two treatments by introducing two different ways to present the LLM-powered analysis in AI-assisted decision making:
\begin{itemize}
    \item \textsc{Control}: In this treatment, we did not incorporate LLM-powered analysis into the AI-assisted decision making process. Participants assigned to this treatment were asked to make decisions with the assistance provided by the AI model alone, without any additional analysis from the LLM.

    \item \textsc{Sequential} (\textsc{Seq}): In this treatment, participants started working on the decision making task seeing only the task instance and the AI model's recommendation without receiving any LLM-generated analyses. Then, we told participants that an LLM had analyzed how different features of the task instance may contribute to the AI model's recommendation on this task. Participants were required to interact with the LLM through a designated interface where, in each turn, the LLM's analysis about one task feature's contribution to the AI recommendation would be \textit{randomly} sampled from the generated set and presented to the participant. The participant could respond to the analysis by indicating whether they agreed or disagreed with it. The participant must interact with the LLM for \textit{at least} $X$ rounds where $X$ is randomly sampled between 1 and 3 for each task. After meeting the minimum interaction requirement, %they could choose to continue until all analyses were provided or to make a final decision. 
    participants could continue to review the LLM-powered analysis on more features, or they could stop the interaction and make their final decisions at any point that they wish. Figure~\ref{interface_seq} shows an example of the task interface used in this treatment.
    \item \textsc{All}: In this treatment, we presented all the LLM-powered analyses for each one of the task features to participants at once, along with the task instance and the AI model's decision recommendation. After reviewing all this information, participants made their final decisions. Figure~\ref{interface_all} shows an example of the task interface used in this treatment.
\end{itemize}

\subsection{Experimental Procedure}
\label{exp_procedure}
%We posted our experiment on Prolific to recruit participants.
Our experiment was conducted on Prolific. 
Upon the arrival of a participant, we randomly assigned them to one of the two types of tasks and one of the three treatments. In the experiment, participants were asked to first fill out an initial survey to report their demographic information and knowledge of AI models. Then, they started the formal experiment by completing a tutorial that described the decision making task they needed to work on. To help participants get familiar with the decision making task, we set up a training stage in which participants completed five decision making tasks independently without seeing the AI model's recommendation or any LLM-powered analyses.
%\zx{is it AI model prediction? make sure you don't confuse readers with AI model prediction and LLM analysis}. 
During these training tasks, we immediately provided participants with the correct answer at the end of each task. After completing the training tasks, participants moved on to the real tasks. In the real tasks, each participant was asked to complete a total of 15 decision making tasks in the assigned treatment. 
%Finally, participants were required to complete an exit survey to report their cognitive load in completing the task, measured by the NASA Task Load Index (NASA TLX)~\cite{}. 
% \zx{convert to hourly payment?}
We offered a base payment of \$1.20 and a raffle with \$1 bonus if the participant’s accuracy was above 85\%. The experiment was open to U.S.-based workers only, and each worker could only complete the experiment once. Additionally, we included two attention check questions in the experiment where participants were required to select a pre-specified option, and only the data of those subjects who passed both attention checks was considered valid. After filtering out the inattentive participants, for the income prediction task, we obtained data from 134 participants (\textsc{Control}: 41, \textsc{Seq}: 45, \textsc{All}: 48), while for the recidivism prediction task, we obtained data from 150 participants (\textsc{Control}: 49, \textsc{Seq}: 40, \textsc{All}: 61). The median working time for participants was about 8 minutes, which translates to a median hourly payment of \$8.9 per hour. For more details of the experiment and participant demographics, please see the supplemental material. 
% \my{Put the materials to the SM!}

% \my{Be precise. How many participants in total? The precise number per treatment should also be listed!}
%\zx{in each task? how many tasks do you have in total?}\zy{update} 

%refer to Appendix~\ref{app:pilot}.

\begin{figure*}[t]
  \centering
  \subfloat[Accuracy]{\includegraphics[width=0.33\textwidth]{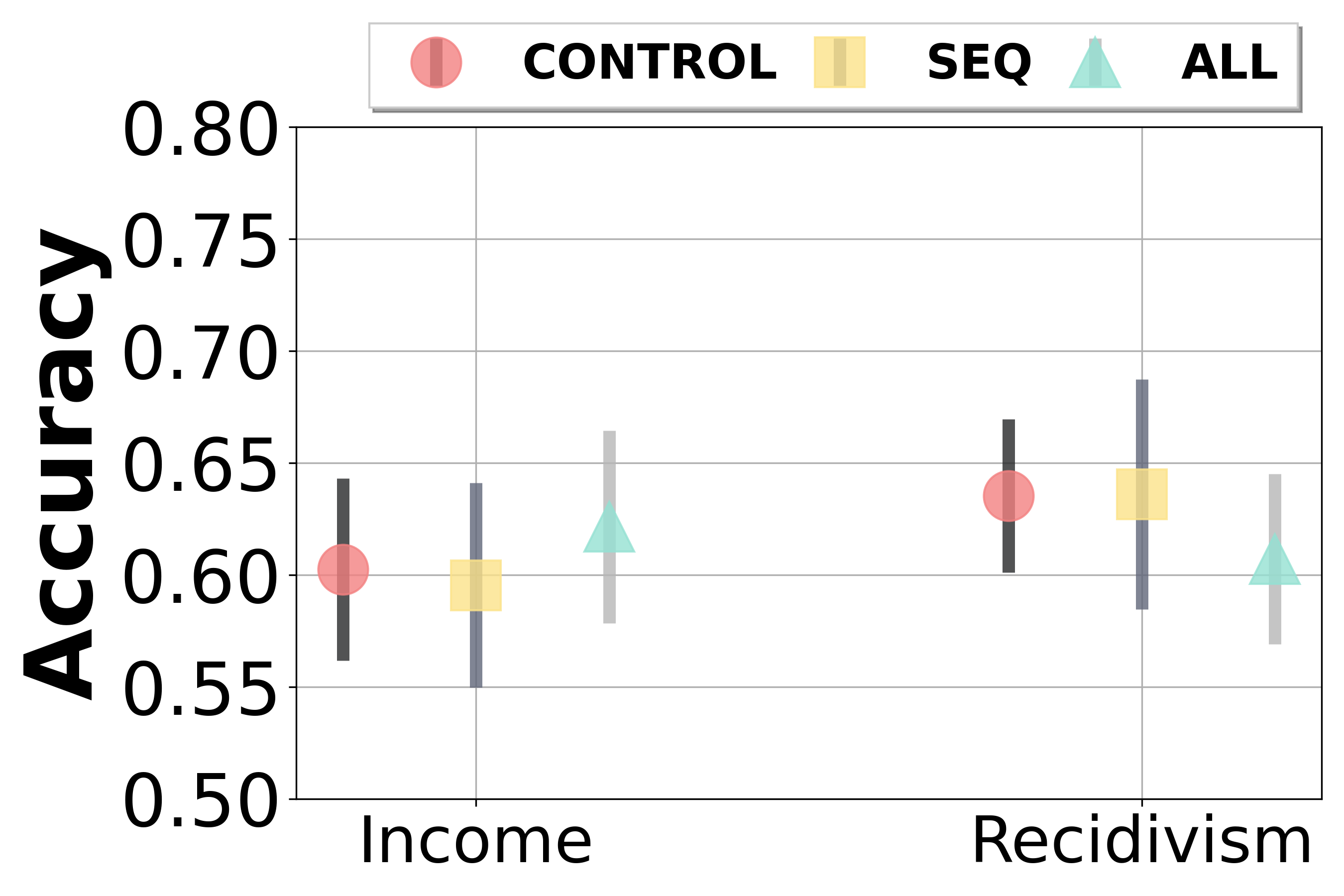}\label{fig:acc_pilot}}
  \hfill
  \subfloat[Overreliance]{\includegraphics[width=0.33\textwidth]{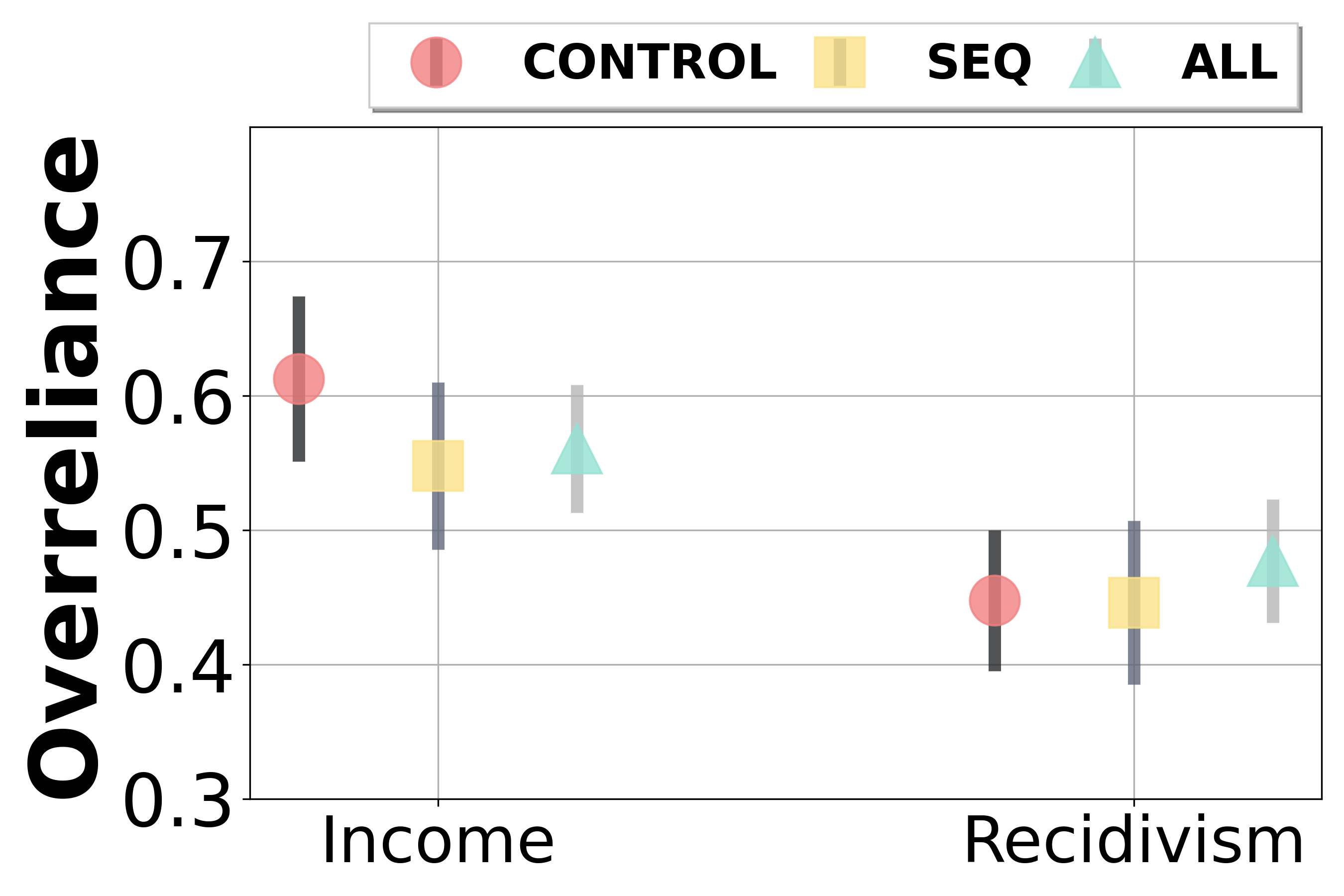}\label{fig:or_pilot}}
  \hfill
  \subfloat[Underreliance]{\includegraphics[width=0.33\textwidth]{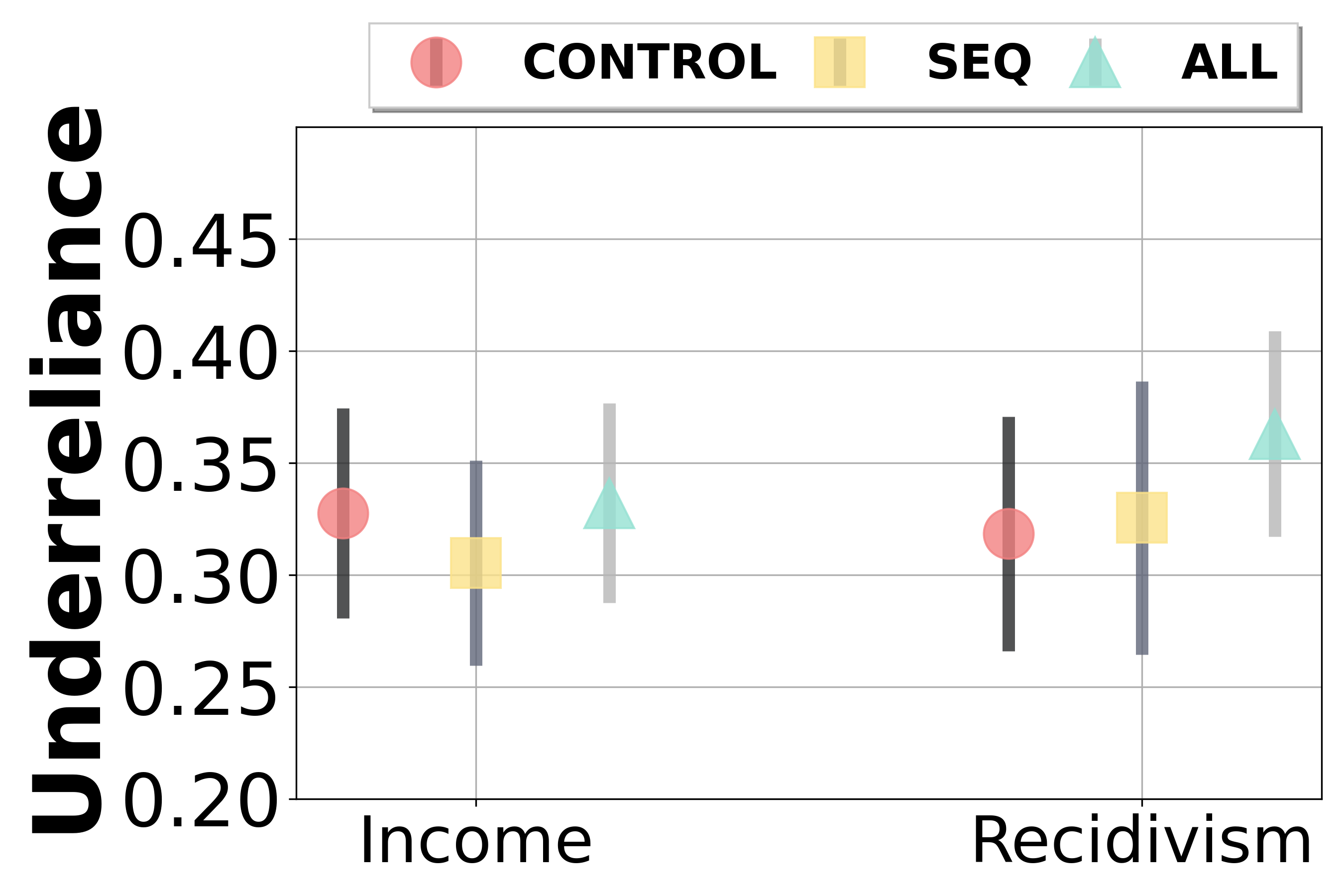}\label{fig:ut_pilot}}
  % \vspace{-10pt}
  \caption{ Comparing the average decision accuracy, overreliance, and underreliance on the AI model for participants across the \textsc{Control}, \textsc{Seq}, and \textsc{All} treatments, for both the income prediction and the recidivism prediction tasks. Error bars represent the 95\% confidence intervals of the mean values. 
  }
\label{pilot_fig}
% \vspace{-10pt}
\end{figure*}

\subsection{Experimental Results}
Following previous work~\cite{zhang2020effect,lai2020chicago}, we used participants' \textit{decision accuracy} to measure the human-AI team performance in decision making, while \textit{underreliance} and \textit{overreliance} were used to quantify the degree to which participants' reliance on the AI model is appropriate. %to measure the human-AI team performance, and the human's appropriate reliance level on AI models. 
Overreliance refers to the fraction of tasks in which the participant's decision was the \textit{same} as the AI model's decision among all tasks where the AI model's decision was incorrect. Underreliance refers to the fraction of tasks in which the participant's decision was \textit{different} from the AI model's decision among all tasks where the AI model's decision was correct. \textit{Lower} overreliance and underreliance indicate that participants' reliance on AI is more appropriate.

Figure~\ref{pilot_fig} shows the comparisons of participants' decision accuracy, overreliance, and underreliance on the AI model across the three treatments for both the income prediction and the recidivism prediction tasks. 
% \zx{should we describe the trend a bit?}\zy{it seems that we don not observe a consistent trend} 
We found that compared to the \textsc{Control} treatment where participants did not receive any LLM-powered analysis, incorporating LLM-powered analyses in AI-assisted decision making does not appear to significantly change participants' decision accuracy or reliance on AI, no matter how they are presented (i.e., sequentially or concurrently). 
%they are presented sequentially (as done in the \textsc{Seq} treatment) or concurrently (as done in the \textsc{All} treatment).
%result in clear differences in participants' performance and reliance on AI models. 
Our one-way ANOVA %\footnote{Analysis of Variance (ANOVA) is a statistical test for identifying significant differences between group means.} 
test results further confirmed that the differences in accuracy, overreliance, and underreliance across the three treatments are not significant at the level of $p=0.05$ for both types of decision making tasks. 
In other words, the ways that human decision makers interact with the LLM-powered analysis in both %the ways humans interact with LLMs in 
the \textsc{All} and \textsc{Seq} treatments may still be not effective for them to critically reflect on the task and calibrate their reliance on the AI recommendation. %present challenges in highlighting critical information from the abundance of analytical analyses provided by the LLM. 
For example, in the \textsc{All} treatment, the sheer volume of information that people need to process may cause a significant cognitive burden, and make it challenging for people to grasp the essential insights from all the information. Meanwhile, in the \textsc{Seq} treatment, although the LLM-powered analysis is presented sequentially to enable decision makers to digest and reflect on each analysis, the random order in which the analysis is presented may imply a miss of opportunity to help decision makers prioritize the most crucial information needed for correct decisions.

\section{Algorithmic Selection of LLM-Powered Analysis in AI-assisted Decision Making}
\label{algo}
Results of our experimental study suggest that in AI-assisted decision making, although LLMs possess the analytical ability to generate additional information to assist humans, the current ways that humans interact with them are not yet optimal. %\my{``interaction ways'' is weird. Just say ``ways that humans interact with ...''}\zy{updated}\zx{I am wondering if necessary to use human humans / humans for human given there is no other humans in your experiment setting. will it be easier to just call it human?}. 
This suboptimal interaction makes it difficult for humans to effectively utilize the information provided by the LLM, hindering their ability to identify essential insights and make informed decisions. Given these challenges, a natural question arises: \textit{How can we enhance the interaction between humans and the LLM to help humans better utilize the analysis provided by the LLM in AI-assisted decision making?} %maximize the effectiveness of the LLM's analytical capabilities in AI-assisted decision making? 
To answer this question, we propose an algorithmic framework that dynamically and strategically selects the most useful LLM-powered analysis to present to human decision-makers, aiming to help them rely on the AI model more appropriately and make correct decisions. 
%based on critical insights from the LLM.

\subsection{Modeling the Effects of LLM-powered Analysis on Human Decision}

\begin{figure*}[t]
\centering
\includegraphics[width=0.95\linewidth]{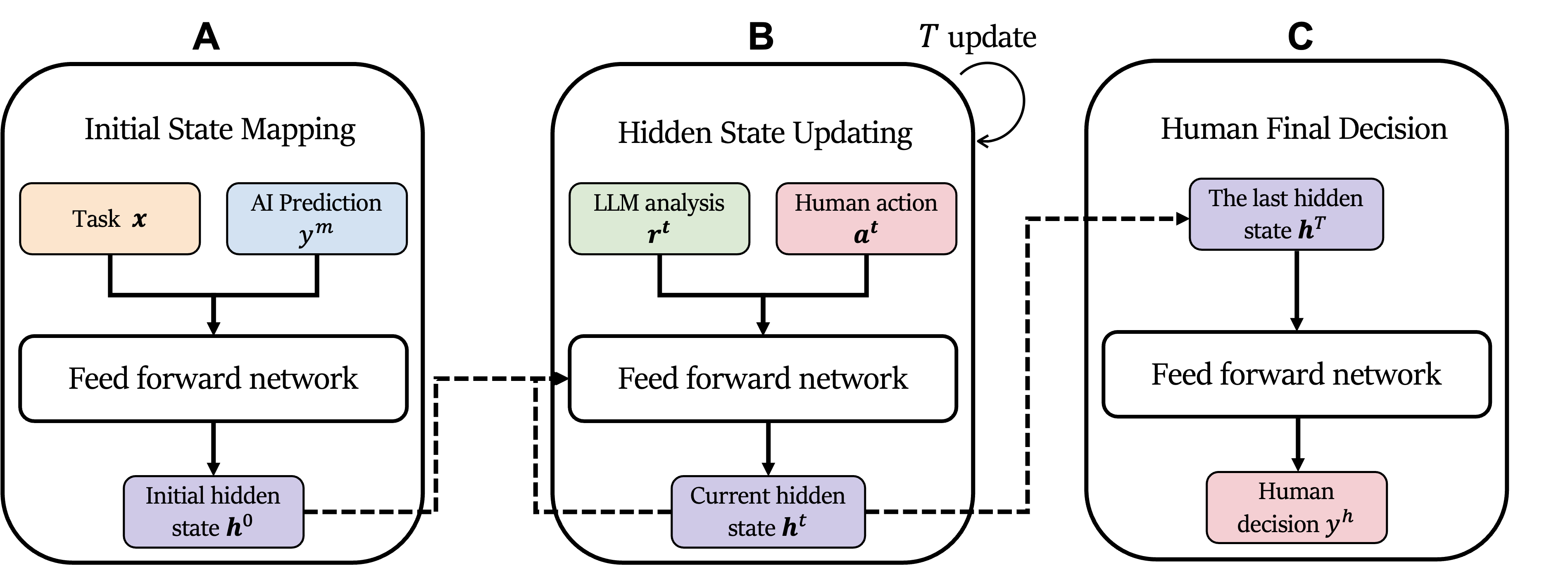}
\caption{Our human behavior model comprises three components: {\bf A}) Initial State Mapping: this component encodes the decision making task and AI recommendation into the human's initial hidden state, which serves as the foundational setup to integrate the task details and initial AI insights into human decision making process. {\bf B}) Hidden State Updating: This component characterizes how the human's hidden state evolves based on the presented LLM-powered analysis and the human's reactions (i.e., whether they agree or disagree with the LLM's analysis). Each update is dependent on the previous hidden state, reflecting the iterative incorporation of new information and human reasoning process into the decision making process. {\bf C}) Final Decision: This component maps the human's latest hidden state to the actual decision made on the task. It translates the cumulative understanding and reasoning process through the hidden states into the human actual decision outcome.}
\label{fig:model}
\vspace{-10pt}
\end{figure*}

To enable the optimal selection of the LLM-powered analysis, we start by quantitatively characterizing how the presentation of LLM-powered analysis impacts humans' decision making in an AI-assisted task. Specifically, consider a human who needs to complete a decision making task with the aid of an AI model. The human is initially provided with the task $\boldsymbol{x} \in \mathcal{X}$ and the AI model's decision recommendation $y^m \in \mathcal{Y}$. Subsequently, the human interacts with the LLM over several rounds to obtain analyses of the task features. In each interaction round $t$ ($1 \leq t \leq T$)\footnote{$T$ is the maximum rounds of interaction occurred, which varies with the specific decision task and may vary across decision makers.}, the human receives a LLM-powered analysis  $\boldsymbol{r}^t \in \mathcal{R}^{t} =
\mathcal{R} \setminus \{\boldsymbol{r}^{1},\ldots,\boldsymbol{r}^{t-1}\}$ from the LLM, where $\mathcal{R}$ is the entire set of analyses generated for the task $\boldsymbol{x}$ across all task features. The human can then reflect on this analysis $\boldsymbol{r}^t$ and indicate their attitude towards it by selecting an option $a^t \in \{\text{agree}, \text{disagree}\}$. After %several 
$T$
rounds of 
interaction with the LLM, the human makes a final decision $y^h \in \mathcal{Y}$. To quantitatively model the effects of these LLM analyses on the human's final decision, we begin by expressing the probability of the final decision given the sequence of interactions as:

\begin{equation}
\begin{split}
        \mathcal{P}(y^{h}|\boldsymbol{x},y^{m},&\boldsymbol{r}^1,a^1,...,\boldsymbol{r}^{T},a^T) = \\ & \int \mathcal{P}(y^{h}|\boldsymbol{h}^{T}) \mathcal{P}(\boldsymbol{h}^{T}|\boldsymbol{x},y^{m},\boldsymbol{r}^1,a^1,...,\boldsymbol{r}^{T},a^T) d\boldsymbol{h}^{T}
\end{split}
\end{equation}
where $\boldsymbol{h}^{T}$ reflects the human's \textit{hidden state} at interaction round $T$. Without loss of generality, we assume that human's hidden state in any round $t$ (i.e., $\boldsymbol{h}^{t}$) is only dependent on the previous hidden state $\boldsymbol{h}^{t-1}$, the LLM-powered analysis presented in the current round (i.e., $\boldsymbol{r}^{t}$), and the human's reaction to this analysis (i.e., $a^{t}$). Thus, we can decompose the above probability as follows:
\begin{small}
\begin{equation}
\begin{split}
      \mathcal{P}&(y^{h}|\boldsymbol{x}, y^{m}, \boldsymbol{r}^1, a^1, \ldots, \boldsymbol{r}^{T}, a^T) = \\&
   \int \underbrace{\mathcal{P}(\boldsymbol{h}^0 | \boldsymbol{x}, y^m)}_{\text{Initial State Mapping}} \underbrace{ \left( \prod_{t=1}^T  \mathcal{P}(\boldsymbol{h}^t | \boldsymbol{h}^{t-1}, \boldsymbol{r}^t, a^t) \right) }_{\text{Hidden State Updating}}
   \underbrace{\mathcal{P}(y^h | \boldsymbol{h}^T)}_{\text{Human Final Decision}}  \, d\boldsymbol{h}^0 \cdots d\boldsymbol{h}^T 
\end{split}
\end{equation}
\end{small}

% \my{Can we put the integral in one line, and put the product as the middle term?}
\noindent Based on this decomposition, our behavior model characterizing how the human's decision is influenced by the LLM-powered analysis consists of three components (see Figure~\ref{fig:model} for a graphical illustration): \\
\noindent \textbf{1. Initial State Mapping}: This component captures the human decision maker's initial hidden state $\boldsymbol{h}^{0}$, before they receive any analysis from the LLM. As shown in Figure~\ref{fig:model}A, we assume that the initial hidden state $\boldsymbol{h}^{0}$ is only influenced by the task instance $\boldsymbol{x}$ and the AI model's recommendation $y^{m}$, and a model parameterized by $\boldsymbol{\theta}_{\text{init}}$ can be learned to characterize the conditional probability distribution of $\boldsymbol{h}^{0}$:
\begin{equation}
    \boldsymbol{h}^{0} \sim \mathcal{P}(\boldsymbol{h}_0 | \boldsymbol{x}, y^m; \boldsymbol{\theta}_{\text{init}}) 
\end{equation}
\noindent \textbf{2. Hidden State Updating}:  This component characterizes how the human decision maker's hidden state evolves over time as they interact with the LLM, i.e., seeing the LLM-powered analysis in each interaction round, for which they may or may not agree with. As shown in Figure~\ref{fig:model}B, the hidden state $\boldsymbol{h}^{t}$ in the $t$-th round is decided by the previous hidden state $\boldsymbol{h}^{t-1}$, the LLM-powered analysis presented in the current round $\boldsymbol{r}^{t}$, and the human's reaction to it $a^{t}$. A model parameterized by $\boldsymbol{\theta}_{\text{update}}$ can be learned to characterize the conditional probability distribution of $\boldsymbol{h}^{t}$:
\begin{equation}
    \boldsymbol{h}^{t} \sim \mathcal{P}(\boldsymbol{h}^t | \boldsymbol{h}^{t-1}, \boldsymbol{r}^t,a^t;\boldsymbol{\theta}_{\text{update}})
\end{equation}
Note that the current hidden state $\boldsymbol{h}^t$ encapsulates the cumulative information gathered through all previous human interactions with the LLM. It achieves this by iteratively encoding the LLM's analysis $\boldsymbol{r}$ and the human reasoning processes (as indicated by human reactions to LLM's analysis $\boldsymbol{a}$) to update the hidden state. Each iteration integrates new insights from the latest LLM analysis and human responses to reflect humans' evolving understanding of the decision making task. 
\\%the evolving understanding of human decision making process.\\
\noindent \textbf{3. Final Decision}: This component maps the human decision maker's last hidden state at the end of the interaction to the final decision they make on the task. As shown in Figure~\ref{fig:model}C, the final decision $y^h$ is only decided by the last hidden state $\boldsymbol{h}^{T}$, and a model parameterized by $\boldsymbol{\theta}_{\text{decision}}$ is used to characterize the conditional probability distribution of $y^h$:
\begin{equation}
    y^{h} \sim \mathcal{P}(y^h | \boldsymbol{h}^T;\boldsymbol{\theta_{\text{decision}}})
\end{equation}
 With a set of human behavior data indicating how humans react to LLM-powered analysis and then make decisions, i.e., $\mathcal{D} = \{\boldsymbol{x}_{i},y^{m}_{i},\{\boldsymbol{r}_{i}^{t},a_{i}^{t}\}_{t=1}^{T}, y^{h}_i\}_{i=1}^{N}$, we can use maximum likelihood estimation to learn the behavior model parameters $\Theta = \{\boldsymbol{\theta}_{\text{init}},\boldsymbol{\theta}_{\text{update}},\boldsymbol{\theta}_{\text{decision}}\}$. 

\subsection{Selecting the LLM-powered Analysis}

% \begin{figure}[htbp]
% \centering
% \includegraphics[width=\linewidth]{figures/flow.003.png}
% \caption{LLM-powered Analysis Selection.}
% \label{fig:workflow}
% \vspace{-10pt}
% \end{figure}

Given a learned model $\Theta$ that characterizes the impacts of LLM-powered analyses on humans' decisions, we next explore how to dynamically select the optimal analysis $\boldsymbol{r}^t$ from the set of candidate analysis (i.e., $\mathcal{R}^{t}$) to maximize human's appropriate reliance on AI models in AI-assisted decision making. To achieve this, we first need to measure the reliability of the AI model's prediction $y^m$ on each task instance $\boldsymbol{x}$. Recent work~\cite{steyvers2022bayesian,kerrigan2021combining} has proposed methods to leverage the complementary strengths of humans and AI in decision making tasks by combining the human's independent decisions and an AI model's decision recommendations intelligently (e.g., using a Bayesian modeling framework), which often yields more accurate decisions than those made by either the human or the AI model alone. Specifically, given the human's independent decision $y^{h}_{\text{independent}}$, the AI model's recommendation $y^m$, and the task instance $\boldsymbol{x}$, these methods learn models to combine $y^{h}_{\text{independent}}$ and $y^m$ to produce a combined result:
\begin{equation}
    y_{\text{combine}} = \text{CombineModel}(y^{h}_{\text{independent}}, y^m, \boldsymbol{x})
\end{equation}
In this study, we adopted the human-AI combination method proposed by \citeauthor{kerrigan2021combining}~\shortcite{kerrigan2021combining} to obtain $y_{\text{combine}}$. Since the accuracy of $y_{\text{combine}}$ was shown to be higher than either $y_{\text{independent}}^h$ and $y^{m}$, we treated $y_{\text{combine}}$ as
the ``target'' decision and we selected the LLM-powered analysis  $\boldsymbol{r}^t$ in a way to nudge humans into making this target decision\footnote{We evaluated various human-AI combination models, and our results showed that the method proposed by~\citeauthor{kerrigan2021combining}~\shortcite{kerrigan2021combining} generally resulted in combined decisions that outperform AI solo and independent human decisions, as well as other combination methods. The evaluation details can be found in the supplementary material.}. In other words, when $y_{\text{combine}}=y^{m}$, we selected a LLM-powered analysis to nudge humans towards relying on the AI recommendation; otherwise, we nudged humans towards not relying on the AI recommendation. 

To effectively nudge the human towards making the target decision $y_{\text{combine}}$, we first define an \textit{immediate utility function} for evaluating the selection of an analysis %$\boldsymbol{r}^t$
$\boldsymbol{r}$ in round $t$ given the human's hidden state by the end of the previous round is $\boldsymbol{h}^{t-1}$. Since our goal is to maximize the probability that the final decision made by the human aligns with the target decision, the utility function $U(\cdot)$ is defined as:
\begin{tiny}
\begin{equation}
\begin{split}
&U(y_{\text{combine}}|
\boldsymbol{r},
\boldsymbol{h}^{t-1}) = \\&
      \log \frac{ \mathbb{E}_{a^t \sim \text{Bern}(0.5)} [\int \mathcal{P}(\boldsymbol{h}^t | \boldsymbol{h}^{t-1}, %\boldsymbol{r}^t,
      \boldsymbol{r}, 
      a^t; \boldsymbol{\theta}_{\text{update}}) \mathcal{P}(y^h = y_{\text{combine}} | \boldsymbol{h}^t; \boldsymbol{\theta}_{\text{decision}}) d\boldsymbol{h}^t]}{\mathcal{P}(y^h = y_{\text{combine}} | \boldsymbol{h}^{t-1}; \boldsymbol{\theta}_{\text{decision}})}
\end{split}
\end{equation}
\end{tiny} 

\noindent 
Here, inside the log term, the numerator represents the probability for the human to make the target decision $y_{\text{combine}}$ if they see the analysis %$\boldsymbol{r}^{t}$ 
$\boldsymbol{r}$
in round $t$, and are asked to \textit{immediately} make a decision by the end of round $t$. 
Since the human's reaction $a^t$ to the analysis %$\boldsymbol{r}^{t}$ 
$\boldsymbol{r}$
is unknown at this point, %the analysis  $\boldsymbol{r}^{t}$, 
when computing this probability, 
we assume that the human is equally likely to agree or disagree with the analysis, %action follows a Bernoulli distribution with a probability of 0.5, 
i.e., $a^t \sim \text{Bern}(0.5)$.
Moreover, the denominator represents the probability that the human would have made the target decision at the end of the round $t-1$.  
%This function measures %whether presenting 
%the \textit{ gain} in the probability that the human will make the target decision $y_{\text{combine}}$ %if we choose to present the LLM-powered 
%after seeing the analysis  $\boldsymbol{r}^{t}$ in round $t$ and being asked to make a decision immediately (i.e., the numerator inside the log term), % whether 
%would increase 
% is larger 
%compared to that the human will make the target decision by the end of round $t-1$ %that in the case 
%where we choose to no longer present any additional LLM-powered analysis to the human 
%(i.e., the denominator inside the log term).} 
%help to bring the predicted human decision $y^h$ closer to the targeted decision $y_{\text{combine}}$. 
Intuitively, $U(y_{\text{combine}}|\boldsymbol{r},\boldsymbol{h}^{t-1})$ reflects the immediate \textit{probability gain} for the human to select the target decision after they are  presented with analysis $\boldsymbol{r}$ in round $t$---when $U(y_{\text{combine}}|%\boldsymbol{r}^{t},
\boldsymbol{r},
\boldsymbol{h}^{t-1})>0$, it means that presenting %$\boldsymbol{r}^{t}$ 
$\boldsymbol{r}$
to the human in round $t$ \textit{increases} their chance of selecting the target decision compared to that at the end of previous round; otherwise, the human's probability of selecting the target decision would decrease or remain the same.

Note that when we need to select the analysis to be presented in round $t$, %the LLM-powered analysis $\boldsymbol{r}^{t}$ in the $t$-th round, 
instead of knowing the human's precise hidden state $\boldsymbol{h}^{t-1}$ at the end of the previous round, we can only recursively estimate a \textit{distribution} of this hidden state using the learned model $\Theta$ and the history of past interactions $\{\boldsymbol{x},y^{m},\{\boldsymbol{r}^{k},a^{k}\}_{k=1}^{t-1}\}$. We denote this distribution of the human's hidden state prior to round $t$ as the ``\textit{state belief}''  $\mathcal{B}(t)$:
\begin{equation}
        \mathcal{B}(t) \propto \mathbb{E}_{\boldsymbol{h}^{t-2} \sim \mathcal{B}(t-1)} \left[ \mathcal{P}(\boldsymbol{h}^{t-1} | \boldsymbol{h}^{t-2}, \boldsymbol{r}^{t-1}, a^{t-1}; \boldsymbol{\theta}_{\text{update}}) \right] \;\;\; \forall t \ge 2
        \label{eqn:beliefUpdate}
\end{equation}
\noindent and $ \mathcal{B}(1) = \mathcal{P}(\boldsymbol{h}^0 | \boldsymbol{x}, y^m; \boldsymbol{\theta}_{\text{init}})$. Thus, given a state belief $\mathcal{B}(t)$, the \textit{expected immediate utility} for selecting the analysis  %$\boldsymbol{r}^{t}$ 
$\boldsymbol{r}$ 
in round $t$ is defined as $\rho(\mathcal{B}(t), y_{\text{combine}},%\boldsymbol{r}^{t}
\boldsymbol{r}) = \mathbb{E}_{\boldsymbol{h}^{t-1} \sim \mathcal{B}(t)}[U(y_{\text{combine}}|
%\boldsymbol{r}^{t},
\boldsymbol{r},
\boldsymbol{h}^{t-1})]$, which represents the \textit{expected probability gain} for the human to  select the target decision after they are presented with the analysis $\boldsymbol{r}$ in round $t$ and are asked to immediately make a decision by the end of round $t$. 

However, note that the human does not have to immediately make a decision by the end of round $t$%right after seeing $\boldsymbol{r}^{t}$
---instead, we could choose to present more LLM-powered analyses to the human if they can help further increase the human's probability of selecting the target decision $y_{\text{combine}}$. Therefore, to determine the optimal analysis  %$\boldsymbol{r}^{t\star}$
that maximizes the ultimate probability for humans to select $y_{\text{combine}}$, 
%expected utility over time, 
we define a value function $V$ to represent the \textit{maximum expected overall utility} that is achievable from the current state belief $B(t)$ given the set of remaining analyses $\mathcal{R}^{t}$:
\begin{small}
\begin{align}
V(&\mathcal{B}(t),  \mathcal{R}^{t}, y_{\text{combine}}) = \max_{%\boldsymbol{r}^t 
\boldsymbol{r}
\in \mathcal{R}^{t}} g(\mathcal{B}(t), \boldsymbol{r}, %\boldsymbol{r}^{t},
y_{\text{combine}})%\nonumber
\\
g(&\mathcal{B}(t),  \boldsymbol{r},%\boldsymbol{r}^{t},
 y_{\text{combine}}) = \nonumber  \\ &\underbrace{\rho(\mathcal{B}(t), y_{\text{combine}},%\boldsymbol{r}^{t}
\boldsymbol{r}
)}_{\text{expected immediate utility}} \nonumber \quad + \underbrace{V(\mathbb{E}_{a^t \sim \text{Bern}(0.5)}[\mathcal{B}(t+1)], \mathcal{R}^{t} \setminus 
%\{\boldsymbol{r}^t\},
\{\boldsymbol{r}\},
y_{\text{combine}})}_{\text{maximum expected future utility}} \nonumber
\end{align}
\end{small}

%where $\gamma$ is the discount factor\footnote{$\gamma$ is set as $1$ in this study.}. 

\noindent In this definition, $g(\mathcal{B}(t),
\boldsymbol{r},
%\boldsymbol{r}^{t},
y_{\text{combine}})$ represents the \textit{expected overall utility} %(i.e., the maximum expected probability gain for the human to select the target decision $y_{\text{combine}}$) 
that is achievable from round $t$ onward when the state belief prior to round $t$ is $\mathcal{B}(t)$ and the analysis $\boldsymbol{r}$ is presented to the human in round $t$. It is composed of two parts. The first part is the expected immediate utility $\rho(\mathcal{B}(t), y_{\text{combine}},%\boldsymbol{r}^{t})
\boldsymbol{r})
$, which represents the \textit{immediate probability gain} for the human to select the target decision in the $t$-th round after %$\boldsymbol{r}^{t}$ 
$\boldsymbol{r}$ 
is presented.
The second part is the maximum expected future utility $V(\mathbb{E}_{a^t \sim \text{Bern}(0.5)}[\mathcal{B}(t+1)], \mathcal{R}^{t} \setminus 
%\{\boldsymbol{r}^t\},
\{\boldsymbol{r}\},
y_{\text{combine}})$, which represents the \textit{maximum future probability gain} for the human to select the target decision in the $(t+1)$-th round and beyond if we continue to present the human with the optimal LLM-powered analyses selected from the set $\mathcal{R}^{t} \setminus 
%\{\boldsymbol{r}^t\}
\{\boldsymbol{r}\}
$, while our state belief prior to the $(t+1)$-th round is $\mathcal{B}(t+1)$. $\mathcal{B}(t+1)$ is updated from $\mathcal{B}(t)$ according to Equation~\ref{eqn:beliefUpdate}, assuming that the human is equally likely to agree or disagree with the analysis  $\boldsymbol{r}$ that is presented in the $t$-th round.
Finally, the optimal LLM-powered analysis %$\boldsymbol{r}^{t\star}
$\boldsymbol{r}^{t}\in\mathcal{R}^{t}$ for round $t$ is selected to
maximize $g(\mathcal{B}(t),\boldsymbol{r}, 
%\boldsymbol{r}^{t},
y_{\text{combine}})$, and the expected overall utility associated with this optimal choice of analysis is denoted as $V(\mathcal{B}(t), \mathcal{R}^{t},y_{\text{combine}})=g(\mathcal{B}(t),\boldsymbol{r}^{t},
y_{\text{combine}})$.

We can iteratively update the value function $V(\mathcal{B}(t), \mathcal{R}^{t},y_{\text{combine}})$ until convergence, which yields the optimal policy $\pi(\mathcal{B}(t),\mathcal{R}^{t},y_{\text{combine}})$ for selecting the optimal analysis %$\boldsymbol{r}^{t\star}$ 
$\boldsymbol{r}^{t}$
to present in the $t$-th round
%that maximizes the expected utility
:
\begin{small}
\begin{equation}
\label{policy}
\begin{split}
    \boldsymbol{r}^{t}
 &=\pi(\mathcal{B}(t), \mathcal{R}^{t},y_{\text{combine}})  \\& =
 \begin{cases} 
\text{Not presenting and stop interaction} \; \;\; \text{if} \; V(\mathcal{B}(t),\mathcal{R}^{t},y_{\text{combine}}) \le 0  \\
\arg\max_{\boldsymbol{r} \in \mathcal{R}^{t}} g(\mathcal{B}(t),\boldsymbol{r},y_{\text{combine}}) \;\;\;\;\;\;\text{otherwise}
\end{cases} 
\end{split}
\end{equation}
\end{small}

\noindent If the value function $V(\mathcal{B}(t),\mathcal{R}^{t},y_{\text{combine}})$ is less than or equal to zero, it indicates that further interaction with the LLM is not expected to %provide positive utility for nudging the human 
increase the chance for the human to make
the target decision. Therefore, we stop presenting LLM analyses and let the human make the final decision. Otherwise, we will present the analysis that maximizes the expected overall utility. 

\section{Evaluation of Algorithmic Framework}
In this section, we explore whether and how our proposed framework, which adaptively presents LLM-powered analysis %based on the estimation of 
by estimating 
the human's hidden state %and our quantitative insights into 
and the effects of LLM-powered analysis on human decisions, can enhance human's decision performance in AI-assisted decision making and calibrate human trust in AI models. %Additionally, we investigate whether the framework promotes more efficient %communication 
%interactions between humans and LLMs.
% \my{Add some transition sentences...}
% \zx{how about ``estimation of human's hidden state and utility of LLM-powered analysis''?}
\subsection{Operationalizing the Algorithmic Framework}
% \my{I wonder if we should put all the details above the the SM...}
We operationalized our proposed algorithmic framework in Section~\ref{algo} in the context of AI-assisted income prediction and recidivism prediction tasks. Specifically, 
we utilized the data collected in Section~\ref{empirical_work} under the \textsc{Seq} treatment to learn parameters $\Theta$ of the human behavior models for both types of decision making tasks. The behavior models are optimized using Adam~\cite{kingma2014adam} with an initial learning rate of $1e-4$ and a batch size of each training iteration of $128$. The number of training epochs is set as $15$. The 5-fold cross validation on the collected data shows that the average accuracy of the learned models in predicting humans' decisions under the assistance of AI recommendations and LLM-powered analysis is $0.74$ for income prediction and $0.71$ for recidivism prediction, respectively. 
To enable the use of the human-AI combination method~\cite{kerrigan2021combining} to infer the target decision for each decision making task, we also conducted a pilot study collecting humans' independent judgments on various income prediction and recidivism prediction tasks. Using this pilot data, we trained two models of humans' independent decision making, which achieved an average accuracy of $0.81$ and $0.84$ for predicting humans' independent judgment in income prediction and recidivism prediction, respectively. 
Finally, we utilized these learned human behavior models and human independent decision making models to dynamically select the LLM-powered analysis for humans in the following study. For more details related to the algorithm setting, please refer to the supplementary material.

\subsection{Experimental Treatments}
In addition to the three baseline treatments discussed in Section~\ref{exp_treatment} (i.e., \textsc{Control}, \textsc{Seq}, \textsc{All}), we introduced two additional experimental treatments for this phase of evaluation:

\begin{itemize} 
% \item \textsc{SHAP}: Participants received both the AI model predictions and the explanations generated by SHAP~\cite{}. After reviewing all the information, participants made their final decisions. (Note: In this treatment, participants did not receive the LLM analysis.)

% \item \textsc{LIME}: Participants received both the AI model predictions and the explanations generated by LIME~\cite{}. After reviewing all the information, participants made their final decisions. (Note: In this treatment, participants did not receive the LLM analysis.)
\item \textsc{Algorithmic (Alg)}:  In this treatment, participants started working on the decision making task seeing only the task instance and the AI model's recommendation without receiving any LLM-generated analysis. Then participants were required to interact with the LLM, where in each turn, the LLM-powered analysis to be presented was selected based on Equation \ref{policy} to nudge the participant towards relying on the AI model's recommendation appropriately. 
%, which evaluates the maximum expected overall utility of selecting different analysis in nudging humans towards relying on the AI model's recommendations appropriately.

\item \textsc{Rank}: This treatment followed the same experimental procedure as the \textsc{Seq} treatment regarding participants' interaction and decision making processes. However, the \textsc{Rank} treatment differed in how the LLM-powered analysis to be presented was selected: We first used the post-hoc XAI method LIME~\cite{10.1145/2939672.2939778} to generate feature importance scores for each task instance and then ranked all task features based on the absolute values of their importance scores. We then selected the LLM-powered analysis to present according to a decreasing order of the absolute importance score of the corresponding feature (instead of in a random order as done in the \textsc{Seq} treatment). 
%LLM analyses for each feature were then presented in this ranked order, unlike in the \textsc{SEQ} treatment, where the LLM  analysis was presented in a random order. 
This treatment is designed to examine whether selecting LLM analysis based on our proposed algorithm—which takes into account potential human reactions to such analyses—can enhance human decision making accuracy compared to selection of LLM analysis that is based solely on heuristic feature importance, should it be available.
% \my{What is LIME? Need to explain...Also, need to motivate why this treatment is necessary.}

\end{itemize}

Finally, as a reference, we also included a \textsc{Human-Solo} treatment where participants completed the decision making tasks on their own \textit{without} receiving either the AI model's recommendation or any LLM-powered analysis.
% \my{You also have a treatment where your proposed algorithm is used, right...}

\begin{figure*}[t]
  \centering
  \subfloat[Accuracy]{\includegraphics[width=0.33\textwidth]{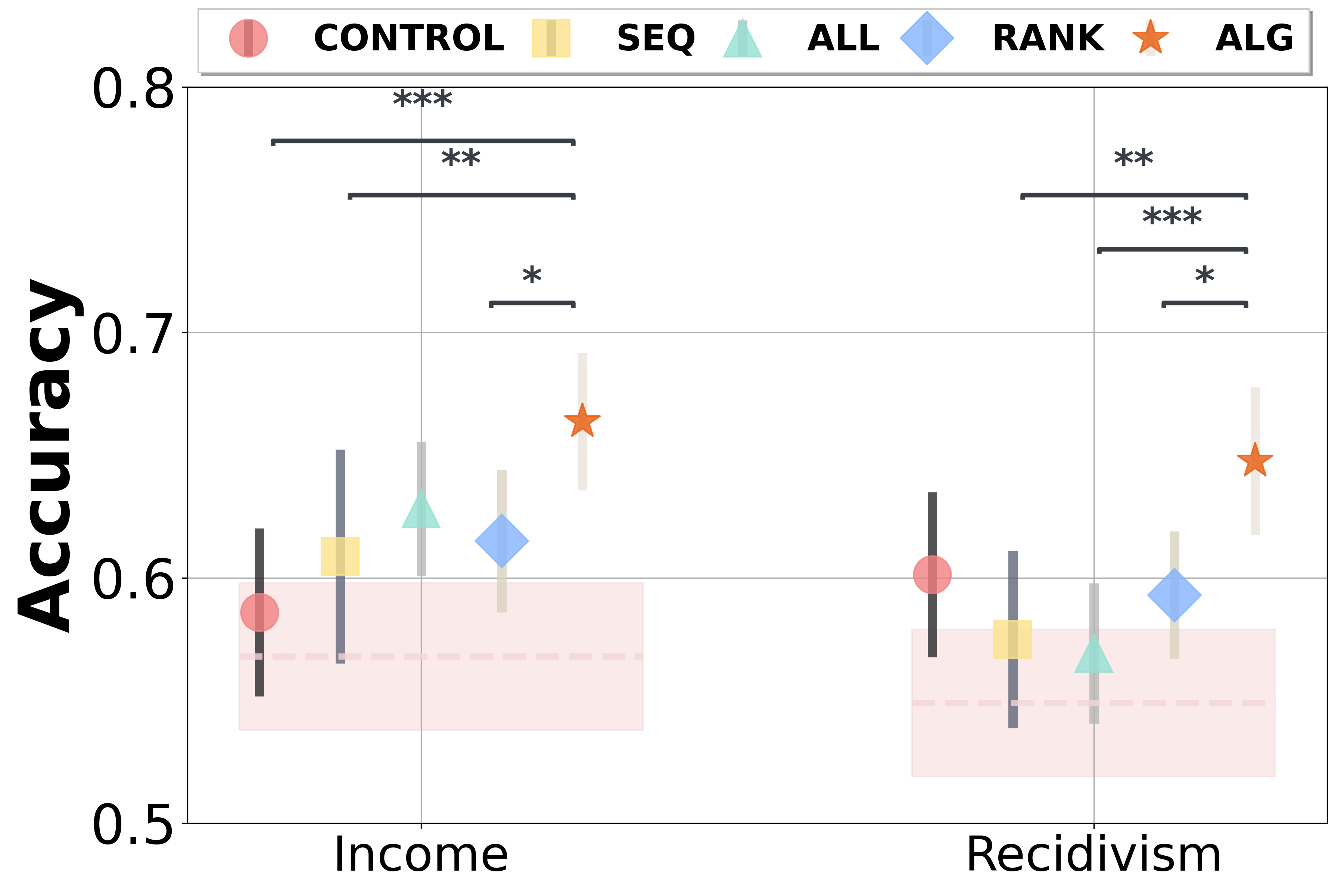}\label{fig:acc}}
  \hfill
  \subfloat[Overreliance]{\includegraphics[width=0.33\textwidth]{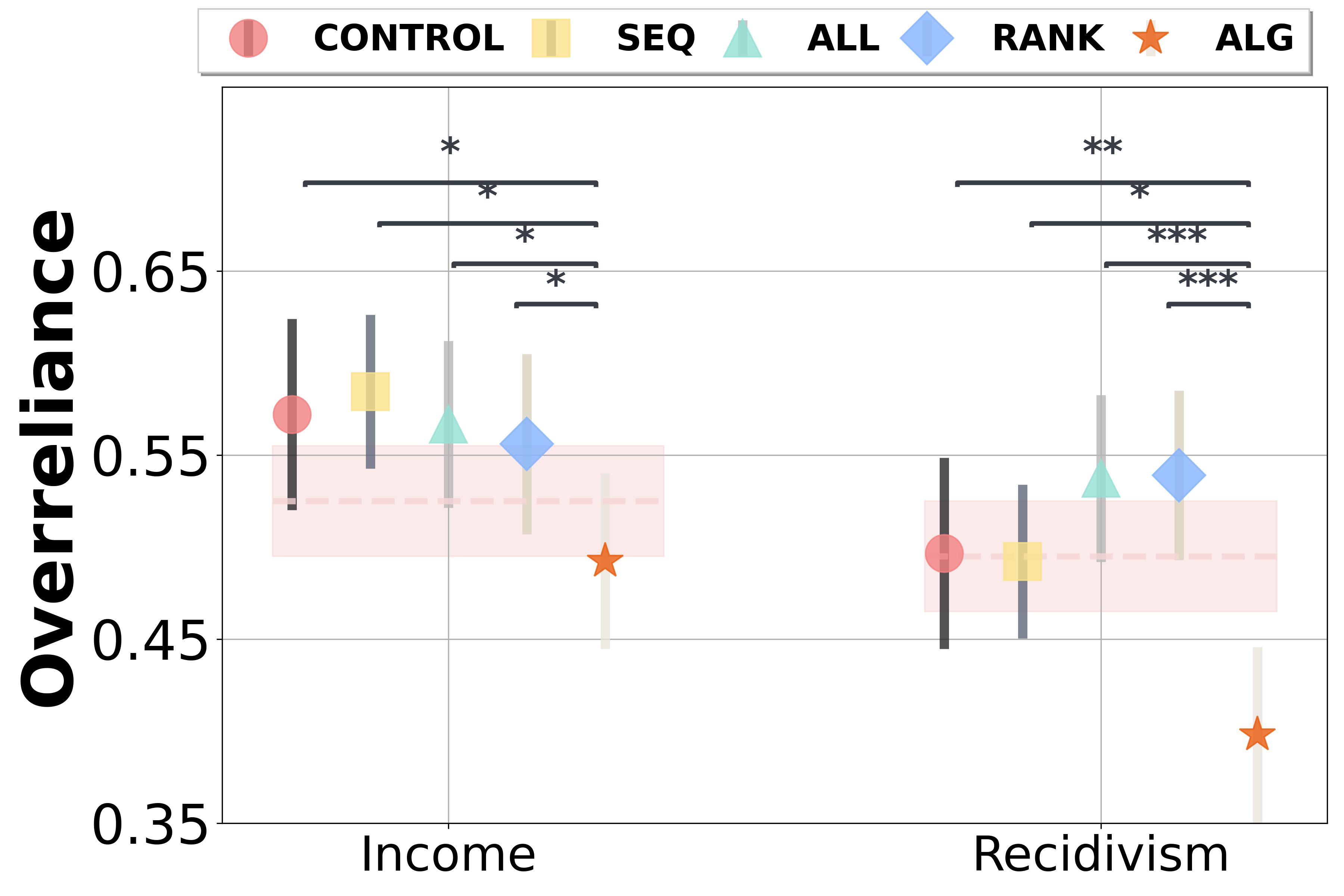}\label{fig:or}}
  \hfill
  \subfloat[Underreliance]{\includegraphics[width=0.33\textwidth]{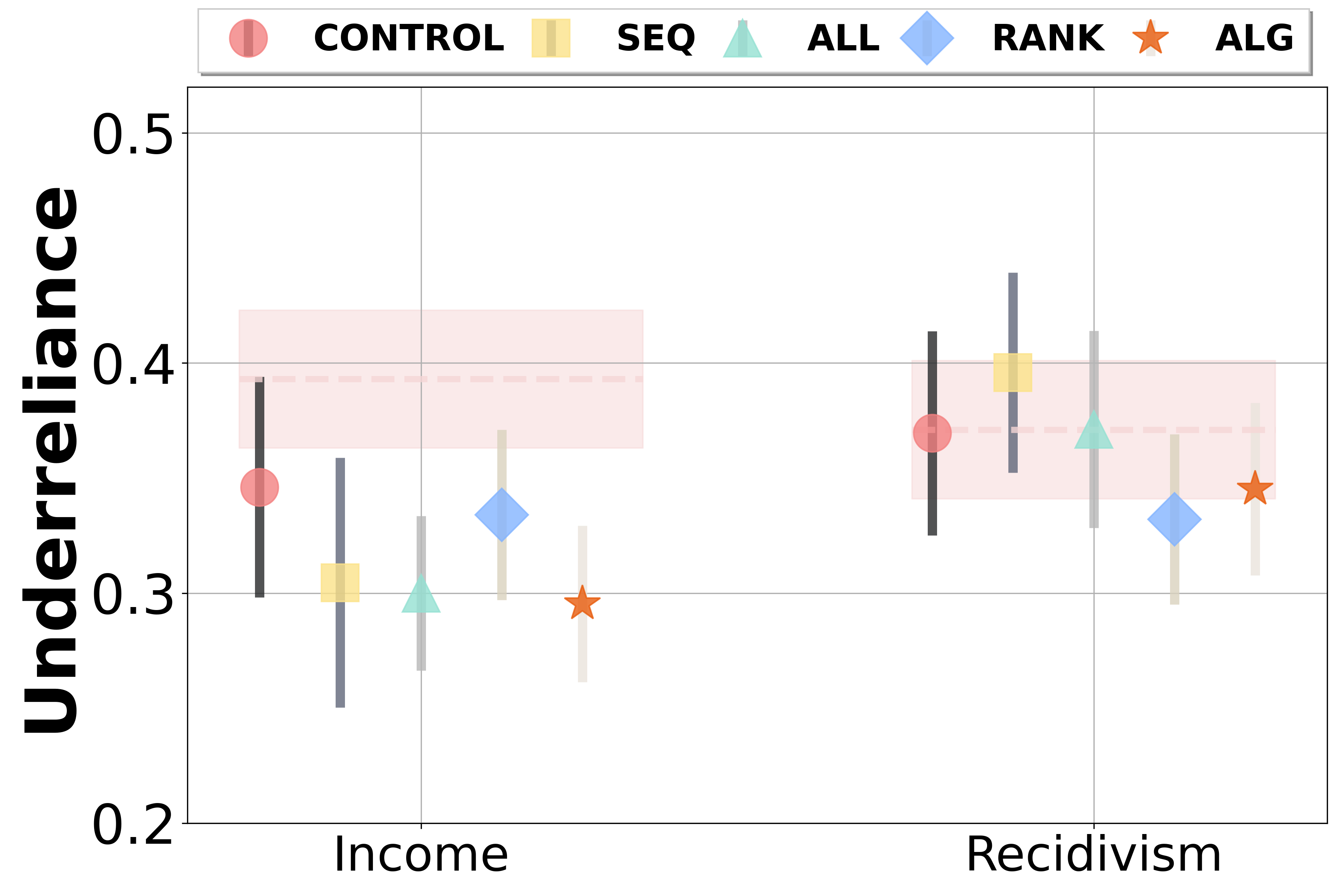}\label{fig:ur}}
  % \vspace{-5pt}
  \caption{ Comparing the participants' average decision accuracy, overreliance, and  underreliance on AI in different treatments for income prediction and recidivism prediction tasks. The pink dashed lines show that for participants in the \textsc{Human-Solo} treatment, (a) the accuracy of their decisions, (b) the frequencies at which their decisions align with AI recommendations (despite not seeing them) when AI recommendations are wrong, and (c) the frequencies at which their decisions differ from AI recommendations (despite not seeing them) when AI recommendations are correct. Error bars (shade) represent the 95\% confidence intervals of the mean values. $\textsuperscript{*}$, $\textsuperscript{**}$, and $\textsuperscript{***}$  denote significance levels of $0.05$, $0.01$, and $0.001$, respectively.}
\label{comp_fig}
% \vspace{-10pt}
\end{figure*}

\subsection{Data Collection}
%\zx{should this be something like study 2?}
Following the experimental procedure described in Section~\ref{exp_procedure}, we again recruited participants from Prolific to complete AI-assisted income prediction and recidivism prediction tasks in the six treatments. 
%once again to collect their decision data under the \textsc{Control}, \textsc{Seq}, \textsc{All}, \textsc{Rank}, and \textsc{Algorithmic} (\textsc{ALG}) 
%treatments.  
For each participant in the income prediction task, we randomly sampled 15 different tasks from a pool of about 500 task instances, which were different from the instances used in either the Section~\ref{empirical_work} study or our pilot study (i.e., these task instances have not be used previously for learning human behavior models or human independent decision models). 
%the Section~\ref{empirical_work} study. 
%\my{What is the Phase 1 study?} to train the human behavior models. \my{I'm confused. I thought your human behavior models are obtained from the data collected in the Seq treatment in Section 3?}\zy{Phase 1 study is just section 3.. will correct the term.} 
Similarly, in the recidivism task, we also randomly sampled 15 different tasks from a pool of about 200 task instances which were different from the task pool used in the Section~\ref{empirical_work} study and our pilot study. We offered a base payment of \$1.20 and a potential bonus of \$1.00 if the participant's decision accuracy was above 85\%. 
We also excluded participants who had previously participated in our study in Section~\ref{empirical_work} or our pilot study from taking this study. 
After filtering out inattentive participants, for the income prediction task, we obtained data from 447 participants, 
while for the recidivism prediction task, we obtained data from 397 participants. 
The median working time of the participants was 9.3 minutes, which translates to a median hourly pay of \$8.3 per hour. 
For more details of the collected data, see the supplementary material.
% \my{Update the median working time, hourly payment, and SM.}

\subsection{Experimental Results}
Below, we analyze whether our proposed algorithmic framework can help decision makers make more accurate decisions, rely on the AI model's decision recommendation more appropriately, and interact with LLM in an efficient manner. 
%promote better utilization of information from LLM-generated analysis by humans, thus yielding improved final decision performance.

\subsubsection{Comparisons of Decision Accuracy}
Figure~\ref{fig:acc} %\ref{fig:or}, and \ref{fig:ur} 
compares the average decision accuracy %, overreliance, and underreliance on AI 
of our participants across treatments. 
%, respectively. %\textsc{Control}, \textsc{Seq}, \textsc{All}, \textsc{Rank}, and \textsc{Alg} treatments, respectively. 
Visually, it appears that participants in the \textsc{Alg} treatment achieve 
the highest decision accuracy among participants in all treatments for both types of tasks.

To examine whether these differences are statistically significant, %with regard to the \textsc{Alg} treatment,
we conducted regression analyses. Specifically, the primary independent variable of the regressions was the treatment participants were assigned to. The dependent variable was the participants' decision accuracy. %, as well as overreliance and underreliance on the AI models. 
To minimize the impact of potential confounding variables, we included a set of covariates in the regression models, such as participants' demographic background (e.g., age, gender, race, and education level), their knowledge of AI models, and the accuracy of the AI recommendation they received in the tasks. Our regression results indicate that our proposed algorithmic framework can significantly improve 
%the overall human-AI team performance 
humans' decision making accuracy 
in both the income prediction and recidivism prediction tasks. Specifically, 
%when examining the final decision accuracy, we found that 
in the income prediction task, participants in the \textsc{Alg} treatment achieved significantly higher accuracy compared to participants in the \textsc{Control} ($p<0.001$), \textsc{Seq} ($p=0.007$), \textsc{Rank} ($p=0.041$) and \textsc{Human-Solo} ($p<0.001$) treatments. Similarly, in the recidivism prediction task, participants in the \textsc{Alg} treatment achieved significantly higher accuracy compared to participants in the \textsc{Seq} ($p=0.006$), \textsc{All} ($p<0.001$), \textsc{Rank} ($p=0.047$) and \textsc{Human-Solo} ($p<0.001$) treatments.

\subsubsection{Comparisons of Appropriate Reliance on AI}

Figures~\ref{fig:or} and \ref{fig:ur} 
compare participants' overreliance and underreliance on AI across treatments, respectively. For participants in the \textsc{Human-Solo} treatment, despite they did not see the AI model's decision recommendations, 
we still computed their hypothetical overreliance and underreliance (i.e., computed as if the participant was presented with the AI recommendation on each task)
to reflect the natural tendency for
%we computed the frequencies at which the participants' own decision happened to align with the AI recommendation when the AI recommendation on a task is wrong to reflect humans' 
participants' independent judgment to agree with an incorrect AI recommendation (Figure~\ref{fig:or})  %Likewise, we computed the frequencies at which these participants' own decision happened to not align with the AI recommendation when the AI recommendation on a task is correct to reflect humans' 
%and their natural tendency to 
or disagree with a correct AI recommendation (Figure~\ref{fig:ur}).  Here, we again see that participants in the \textsc{Alg} treatment almost always achieve the lowest level of overreliance and underreliance on AI among participants in all treatments. Our regression analyses suggest that for participants in the \textsc{Alg} treatment, the decrease in their overreliance on AI is statistically significant compared to participants in other treatments. For example, in the income prediction task, our proposed framework led to participants' significantly decreased overreliance on AI compared to that of participants in the \textsc{Control} ($p=0.028$), \textsc{Seq} ($p=0.011$), \textsc{All} ($p=0.014$), and \textsc{Rank} ($p=0.034$) treatments. Similarly, in the recidivism prediction task, our proposed framework significantly decreased overreliance compared to \textsc{Control} ($p=0.002$), \textsc{Seq} ($p=0.013$), \textsc{All} ($p<0.001$), and \textsc{Rank} ($p<0.001$) treatments, and it even made participants agree with the wrong AI recommendations less than the natural degree of agreement exhibited by participants in the \textsc{Human-Solo} treatment ($p<0.001$).  On the other hand, our regression results suggest that the decrease in participants' underreliance on AI in the \textsc{Alg} treatment was not statistically significant compared to other treatments; the only exception was that on the income prediction task, participants in the \textsc{Alg} treatment disagreed with correct AI recommendations significantly less than the natural degree of disagreement exhibited by participants in the \textsc{Human-Solo} treatment ($p<0.001$). 
%\my{p-values between 0.05 and 0.1 should only claim marginal significance.}

\begin{table}[]
\begin{tabular}{ccc}
\hline
Treatment                     & Income prediction & Recidivism prediction \\ \hline
\textsc{Seq} & 4.88 $\pm$ 1.79         & 4.71 $\pm$ 1.89           \\
\textsc{Rank} & 4.87 $\pm$ 1.03        & 3.72  $\pm$ 1.33          \\
\textsc{Alg} & \textbf{2.99 $\pm$ 1.51}        & \textbf{2.50  $\pm$ 1.19}           \\ \hline
\end{tabular}

\caption{The mean and standard deviation in the round of interactions between participants and the LLM in the \textsc{Seq}, \textsc{Rank}, and \textsc{Alg} treatments in a single decision making task. 
%The \textbf{bold} text indicates that, 
According to the results of the ANOVA test, followed by Tukey's HSD test, the number of interaction rounds in the \textsc{Alg} treatment is significantly lower than the number in the other treatments.}
\label{interaction_round}
\vspace{-20pt}
\end{table}

\begin{figure*}[t]
  \centering
  \subfloat[Accuracy]{\includegraphics[width=0.33\textwidth]{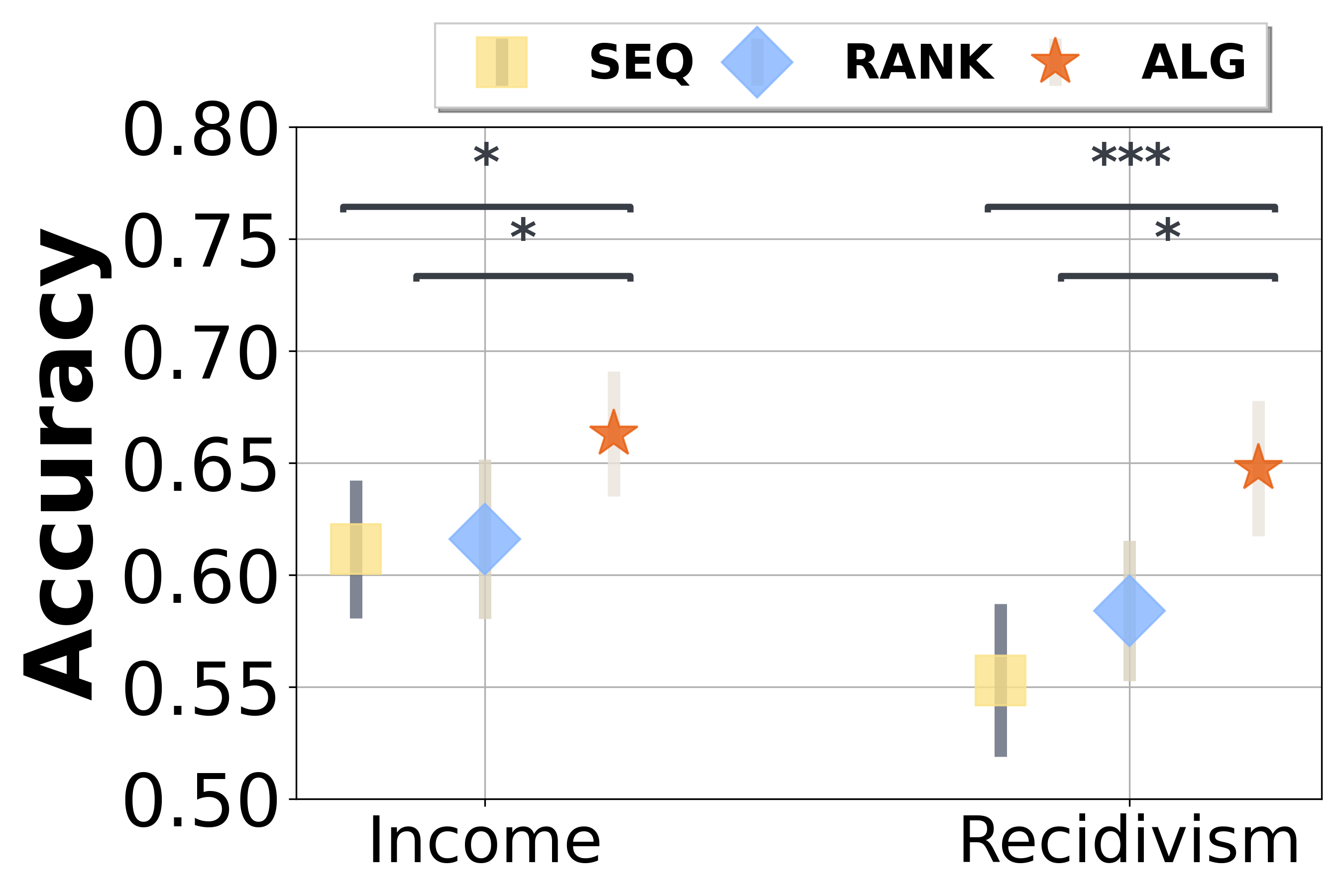}}
  \hfill
  \subfloat[Overreliance]{\includegraphics[width=0.33\textwidth]{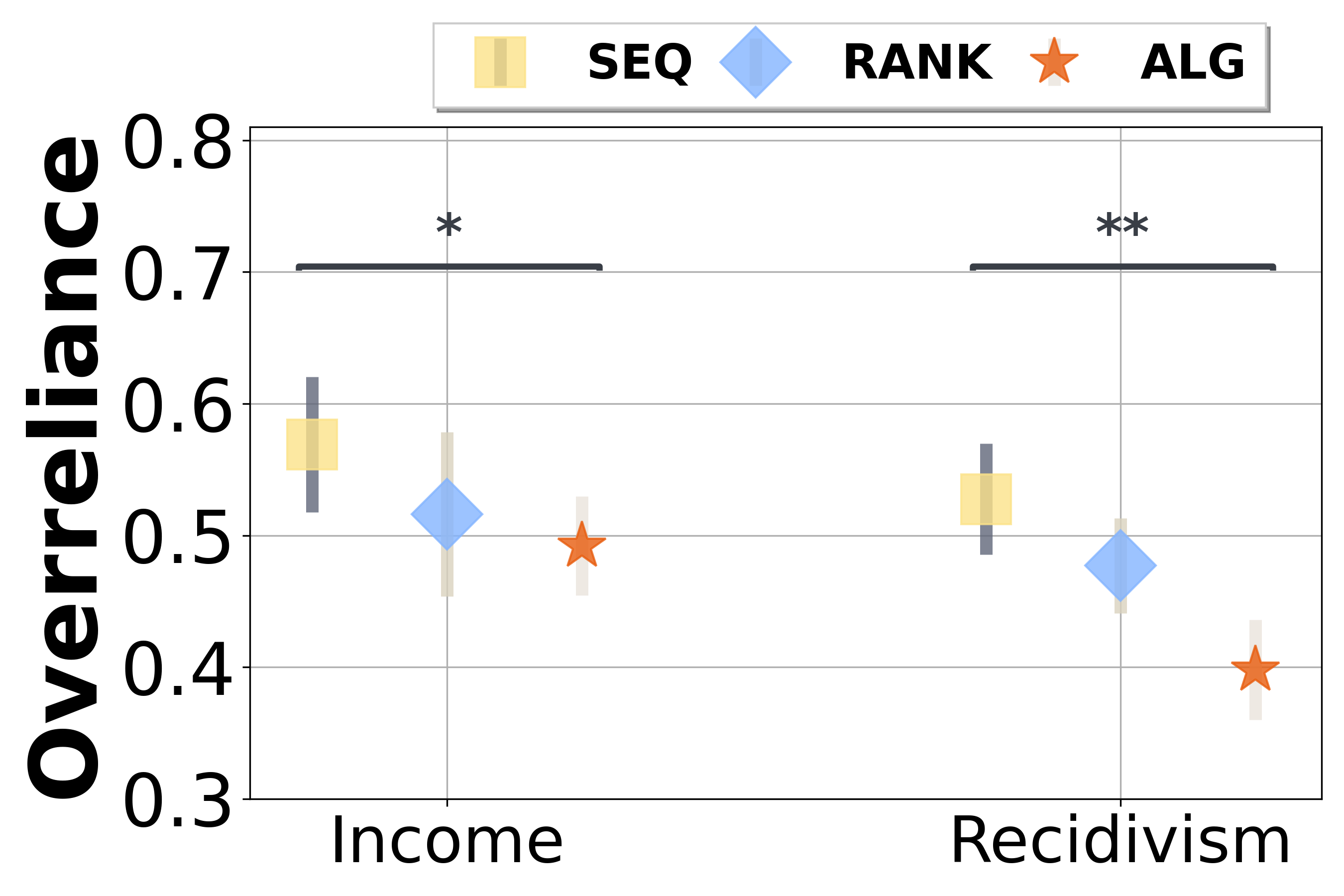}}
  \hfill
  \subfloat[Underreliance]{\includegraphics[width=0.33\textwidth]{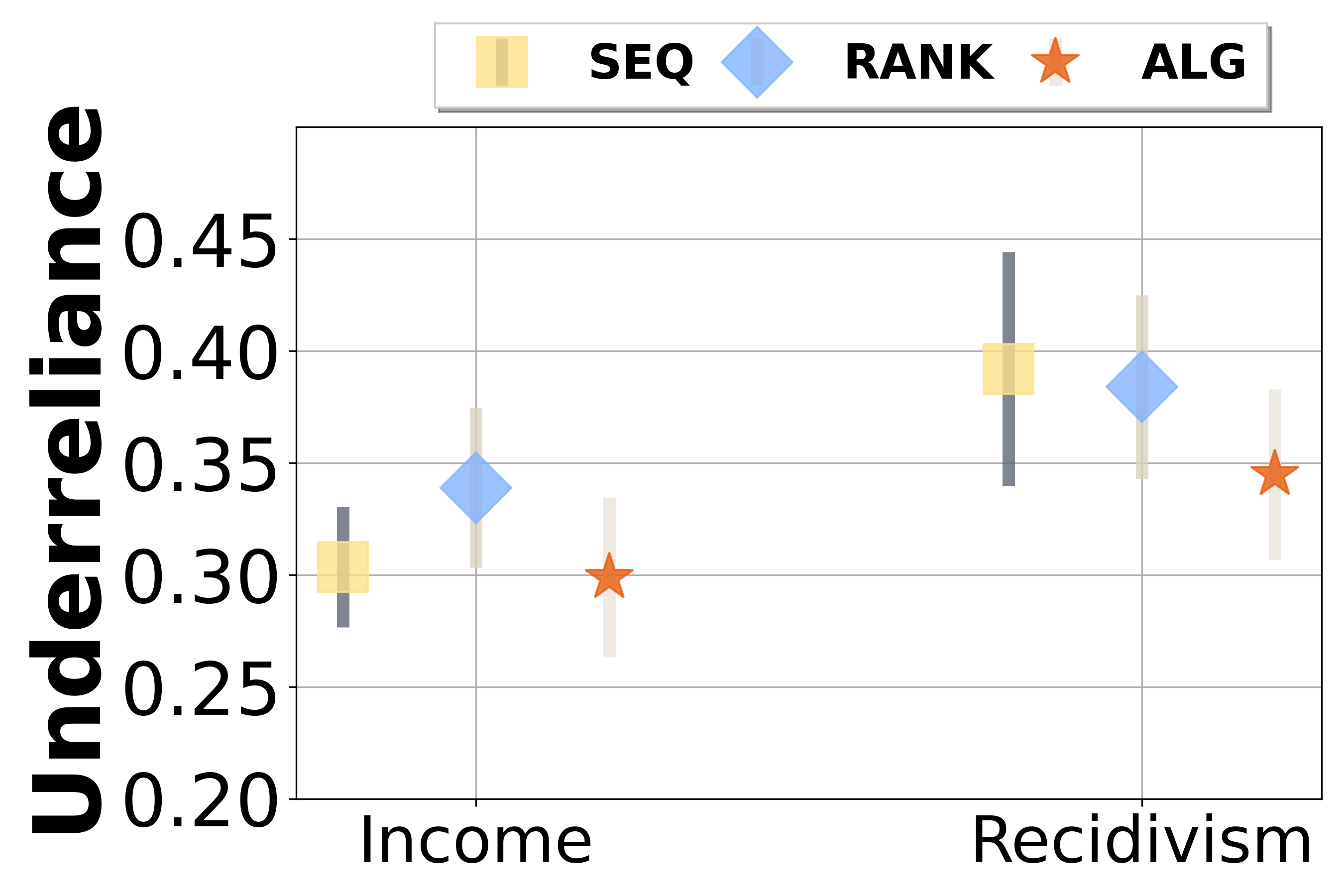}}
  \vspace{-5pt}
  \caption{ Comparing the participants' average decision accuracy, overreliance, and  underreliance on AI in different treatments for income prediction and recidivism prediction tasks, when fixing the number of interaction rounds at the same level. Error bars represent the 95\% confidence intervals of the mean values. $\textsuperscript{*}$, $\textsuperscript{**}$, and $\textsuperscript{***}$  denote significance levels of $0.05$, $0.01$, and $0.001$, respectively.}
\label{comp_fig_same_number}
% \vspace{-10pt}
\end{figure*}

\subsubsection{Comparisons of Efficiency of Interactions}
Lastly, we looked into whether the proposed algorithm helps human decision makers process the most informative information from the LLM-powered analysis in an efficient manner.  First, Table~\ref{interaction_round} compares the average number of interaction rounds between participants and the LLM in the \textsc{Seq}, \textsc{Rank}, and \textsc{Alg} treatments in a single decision making task. Results of ANOVA tests indicate that the number of interaction rounds is significantly different across treatments for both types of decision making tasks ($p<0.001$).
%to determine if there were statistically significant differences in the number of interaction rounds between these treatments. The results showed a significant effect of treatment type on the number of interaction rounds with $p<0.001$ for both income prediction and recidivism prediction tasks. 
We then proceed with post-hoc pairwise comparisons
using Tukey's HSD tests. We found that, for both the income prediction task and the recidivism prediction task, our proposed approach led to significantly fewer rounds of interactions between participants and the LLM compared to the \textsc{Rank} ($p<0.001$ for both tasks) and \textsc{Seq} ($p<0.001$ for both tasks) treatments.  This suggests that our proposed algorithm potentially decreased decision makers' cognitive load and helped them make decisions in a time-efficient manner.

In addition, as our proposed algorithm resulted in the highest decision accuracy among all treatments, it is natural to ask if this increase in accuracy was caused by the decreased number of LLM analysis shown to participants, or by the nature of the LLM analysis selected.
%, while also resulting in higher accuracy. This suggests that our algorithmic selection of LLM-powered analysis can effectively choose the most informative analysis, reducing humans' cognitive load and helping them make better decisions in a 
To gain a deeper understanding on this, we conducted another human-subject experiment with three treatments---\textsc{Seq}, \textsc{Rank}, and \textsc{Alg}---and we  controlled the number of interaction rounds in a decision making task in the \textsc{Seq} or \textsc{Rank} treatments to match that experienced by participants in the \textsc{Alg} treatment\footnote{Based on our results in Table~\ref{interaction_round}, for the income prediction task, we set the number of interaction rounds in a task to be 3. For the recidivism prediction task, we set the number of interaction rounds in a task to be 2 or 3 uniformly randomly.}. %For the \textsc{Seq} and \textsc{Rank} treatments, we set the number of interaction rounds to match those experienced by participants in the \textsc{Alg} treatment. 
For each type of decision making task, we recruited 50 participants for each treatment. Figure~\ref{comp_fig_same_number} compares participants' decision accuracy, overreliance, and underreliance on AI across the three treatments. Again, we found that participants in the \textsc{Alg} treatment achieved significantly higher accuracy compared to 
participants in the 
\textsc{Seq} (income prediction: $p=0.044$, recidivism prediction: $p<0.001$) and \textsc{Rank} (income prediction: $p=0.032$, recidivism prediction: $p=0.012$) treatments.  
%in the income prediction task. In the recidivism prediction task, participants in the \textsc{Alg} treatment also showed significantly higher accuracy than those in the \textsc{Seq} ($p<0.001$) and \textsc{Alg} ($p=0.012$) treatments. 
Moreover, we observed that participants in the \textsc{Alg} treatment significantly decreased their overreliance on AI compared to those in the \textsc{Seq} treatment for both the income prediction task ($p=0.042$) and the recidivism prediction task ($p=0.004$). This means that the proposed algorithm improved the accuracy of participants' decisions and promoted their appropriate reliance on the AI recommendation primarily as it selected the most informative LLM-powered analysis to be presented to people.

\begin{table*}[t]
\centering
\small
\begin{tabular}{c|cc|cc} 
\hline
\multirow{3}{*}{Task} & \multicolumn{4}{c}{Alignment Rate ($\%$)}                                     \\ 
\cline{2-5}
                      & \multicolumn{2}{c|}{Initial Analysis} & \multicolumn{2}{c}{All Analyses}  \\ 
\cdashline{2-5}
                      & AI Correct & AI Incorrect             & AI Correct & AI Incorrect         \\ 
\hline
Income Prediction     & 59         & 25                       & 51         & 21                   \\
Recidivism Prediction & 51        & 38                       & 50         & 37                  \\
\hline
\end{tabular}
\caption{The alignment rate between the LLM analyses and the AI model recommendations when the AI model's decision recommendation is correct or incorrect for the income prediction and recidivism prediction tasks.}
\label{tab:correctness_alignment}
% \vspace{-20pt}
\end{table*}

\begin{table*}[t]
\centering
\small
\begin{tabular}{c|cc|cc} 
\hline
\multirow{3}{*}{Task} & \multicolumn{4}{c}{Alignment Rate ($\%$)}                                     \\ 
\cline{2-5}
                      & \multicolumn{2}{c|}{Initial Analysis} & \multicolumn{2}{c}{All Analyses}  \\ 
\cdashline{2-5}
                      & AI = Target & AI $\neq$ Target            & AI = Target & AI $\neq$ Target        \\ 
\hline
Income Prediction     & 66          & 5                       & 64          & 11                  \\
Recidivism Prediction & 64          & 23                      & 69          & 15                  \\
\hline
\end{tabular}
\caption{The alignment rate between the LLM analyses and the AI model recommendation when the AI model's decision recommendation matches or does not match the target decision for the income prediction and recidivism prediction tasks.}
\label{tab:target_alignment}
% \vspace{-20pt}
\end{table*}

\subsection{Exploratory Analyses}
Finally, to gain deeper insights into why the proposed algorithm effectively nudged decision makers towards making more accurate decisions and relying on AI recommendations more appropriately, we conducted exploratory analyses to understand the nature of the LLM-powered analysis selected by the algorithm.

As the example analysis shown in Table~\ref{tab:llm_rationales}, given a decision task, the LLM typically provides its interpretation on how the value of a task feature influences the prediction---the value of a feature could increase, decrease, or has no influence on the likelihood of a certain prediction. The direction of this suggested influence can either align or not align with the AI model's actual recommendation. For example, on an income prediction task, if the LLM suggests the education level of the person in the task decreases the likelihood of them making over \$50k per year, and the AI model's prediction on the task is indeed ``below \$50k'', then this analysis ``\textit{aligns}'' with the AI prediction (i.e., on this task, the person's education level provides \textit{supporting evidence} to the AI model's recommendation). However, if the AI model's prediction on the task is ``above \$50k'', then this analysis does not align with the AI prediction, and the value of the person's education level provides a \textit{contradictory evidence} to the AI model's recommendation.

For all participants in the \textsc{Alg} treatment of our experiment, we analyzed whether each LLM-powered analysis presented to them on a decision task aligned with the AI model's recommendation on that task. In Table~\ref{tab:correctness_alignment}, we compared the fraction of selected LLM analysis that aligned with AI recommendation for tasks in which the AI recommendation was correct and tasks in which the AI recommendation was wrong, and such comparison was conducted when considering only the first analysis selected by the LLM on each task (see the ``Initial analysis'' column), or considering all analyses selected by the LLM on each task (see the ``All analyses'' column). Clearly, for both income prediction and recidivism prediction tasks, we found that LLM analyses that aligned with the AI recommendation were significantly more likely to be selected on tasks where the AI recommendation was correct than on tasks where the AI recommendation was wrong (proportion tests suggest that $p<0.001$). In other words, when the AI recommendation was correct, our algorithm was more likely to select analysis that provides ``supporting evidence'' to the AI recommendation, while analysis that provides ``contradictory evidence'' was more likely to be selected when the AI recommendation was wrong.

Table~\ref{tab:target_alignment} shows an even larger discrepancy in the alignment rate of the LLM-powered analysis selected when focusing on the comparison between tasks where the target decision was the same as the AI recommendation (hence the algorithm aimed to nudge the participant towards relying on AI), versus tasks where the target decision was different from the AI recommendation (hence the algorithm aimed to nudge the participant towards not relying on AI). This means that the algorithm primarily presents supporting evidence to humans to nudge them to rely on AI, while primarily presents contradictory evidence to humans to nudge them towards not relying on AI. As a qualitative example, in an income prediction task, suppose the AI model predicts the person's income would be above \$50k. If the algorithm aims to increase participants' reliance on this prediction, the top three LLM analysis selected by the algorithm across all decision making tasks are 
``\textit{With an occupation of professional specialty, this might increase the likelihood of making over \$50k per year}'', ``\textit{With the sex being male, this might increase the likelihood of making over \$50k per year}'', and ``\textit{With the work type being in the private sector, this might increase the likelihood of making over \$50k per year}''. 
%``\textit{ With an age of X [X is a value that is below median], this might decrease the likelihood of making over \$50k per year}'', ``\textit{ With the marriage status of divorced, this might decrease the likelihood of making over \$50k per year}'', and ``\textit{ With X [X is a value that is below median] years of education, this might decrease the likelihood of making over \$50k per year}''. 
In contrast, if the algorithm aims to decrease participants' reliance on this prediction, the top three LLM analysis selected by the algorithm across all decision making tasks are 
``\textit{With an age of X [X is a value that is below median], this might decrease the likelihood of making over \$50k per year}'', ``\textit{With a marital status of divorced, this might decrease the likelihood of making over \$50k per year}'', and ``\textit{With X [X is a value that is below median] years of education, this might decrease the likelihood of making over \$50k per year}''.

\section{Discussions}

In this paper, via two phases of study, 
we explore how to effectively incorporate the analytical capabilities of LLMs in AI-assisted decision making  to improve human-AI team performance in the absence of AI explanations. Based on our findings, we discuss the potential societal impacts, design implications, and limitations of our study. 

\subsection{Algorithmic Selection of LLM-powered Analysis Could Be a Double-Edged Sword%Dual Use of LLM-powered Analysis in AI-assisted Decision Making
}
In our study, we seek to enhance human-AI team performance in decision making by selectively and progressively presenting LLM-generated analyses that nudge humans towards making decisions that are considered as optimal by a rational integration of human and machine intelligence. This practice demonstrated potential benefits, such as improving the accuracy of human-AI team's decisions and reducing human overreliance on AI models. As we have shown in our study, the integration of carefully selected LLM-powered analyses in AI-assisted decision making can, under controlled conditions, lead to improved decision making performance by augmenting AI recommendations with detailed task analysis and enabling humans to reflect on the AI recommendations in a structured way. %highlighting overlooked aspects of a problem in areas where human biases or limited information processing capabilities might lead to suboptimal decisions. 

However, our findings also raise concerns about 
%the potential risks of the algorithmic selection of what information humans receive in their decision making. 
the susceptibility of human behavior to algorithmic selection of the information that humans receive in their decision making. 
%such algorithmic manipulation by presenting LLM-powered analysis. 
Despite the apparent benefits, the ease with which human decisions can be influenced by algorithmically selected LLM-powered analysis poses notable risks. Our study reveals that it is relatively straightforward to set a nudge direction that subtly manipulates human decision outcomes. This manipulation, while potentially benign and intended to correct for known biases or decision making flaws, could also be maliciously used by adversarial actors to achieve unethical goals. In the context of recidivism prediction, an example of such misuse could involve an adversarial actor manipulating the human decision making process to be unfairly biased against certain groups~\cite{li2024utilizing}. By setting an unethical nudge goal, the LLM-powered analysis can be algorithmically presented in a manner that selectively emphasizes certain aspects over others. This selective presentation might influence human judicial decisions, nudging them towards more punitive measures for targeted populations, which reinforces existing societal biases and compromises the fairness. 

To counteract the potential adversarial uses of LLM-powered analysis, it is crucial that further research not only focuses on developing and enhancing the capability of AI models to support human decision making, but also on devising strategies to prevent their misuse in manipulating decisions adversely. For instance, to mitigate the risks of adversarial nudges in AI-assisted decision making, strengthening security measures around AI systems like implementing both physical security measures and cybersecurity protocols designed to guard against unauthorized access, hacking, and manipulation is critical. In addition, in our study, the successful nudging of human decisions to improve the human-AI team performance was based upon the accurate modeling of human behavior.  This modeling was fundamentally based on empirical  human-AI interaction data. As such, protecting this data from misuse is crucial. Strict controls must be in place to ensure that only authorized and well-intentioned parties have access to sensitive interaction data, to prevent the misuse of algorithmic nudges.  Finally, implementing continuous monitoring of decisions when humans interact with AI/LLM-powered systems is necessary to detect any unusual patterns in human behavior that may indicate potentially misleading or biased AI information.

\subsection{On Determining Nudging Directions through Combining Human Decisions and AI Recommendations}
In our proposed framework, a key step is to determine the trustworthiness of the AI recommendation and decide whether to present LLM analysis to nudge human decision makers towards relying on the AI recommendation or not. We did so by leveraging the ``human-AI complementarity''---we inferred a ``target decision'' on each decision making task using existing methods (e.g., \cite{kerrigan2021combining}) to combine the predicted human's independent decision on the task and the AI model's recommendation on the task, and nudging human decision makers towards making this target decision. While these combination methods could always be used to generate a target decision, the quality of the target decision---to what extent the target decision is more accurate than both human's independent decision and AI's recommendation and therefore provides useful information on the trustworthiness of the AI recommendation---may vary with many factors. 
%the quality of the ``target'' decision produced by the human-AI combination model is crucial for enhancing human decision making performance, as it sets the goal to nudge human behavior through selective LLM analysis. 
%The previous study~\cite{steyvers2022bayesian} have identified several factors that influence this algorithmic combination result of human and AI decisions. 
For example, the correlation between human and AI decisions was found to be a significant factor that would limit the human-AI complementarity---the more correlated humans' and AI's decisions are, the less likely the combined decision outperforms both human and AI alone~\cite{steyvers2022bayesian}. 
%If humans initially make decisions on certain data points and then use these decisions to train the AI model, 
This implies that if the AI model is trained based on historic decisions made by humans to mimic human decision making,
the algorithmic combination of human and AI decisions may not yield target decisions of significantly higher accuracy.  %To ensure the quality of the target decision generated by the combination model, it is essential to maintain the independence of AI decisions from human decisions and strive for optimal AI model performance. 
Another key influencing factor is the accuracy differences between humans and AI---the larger the accuracy difference, the less likely the combination of human and AI decisions would outperform the decision of the more accurate party~\cite{steyvers2022bayesian,bansal2021does}. Different combination methods may also yield target decisions with varying levels of accuracy, as each method has its own assumptions when modeling human decisions and AI decisions, which may or may not be valid for a specific decision making task.

As the effectiveness of the combination method may vary with many different factors, in practice, given a particular type of decision making task, we recommend first collecting pilot data on human and AI's decisions on this task. This data would enable the comparison of the performance of various combination methods as well as understanding if the combined decisions show true advantages over the independent decisions of either humans' or AI's. If the accuracies of combined decisions are similar to the more accurate party between the human decision maker and the AI model, instead of triggering additional computational cost to compute the combined decisions, one may consider simply nudging the decision maker to always rely on AI (if AI is more accurate than human) or always not rely on AI (if human is more accurate than AI). However, if the combined decisions are more accurate than both humans' and AI's decisions, one should select the combination method that produces the most accurate combined decisions, or even design new combination algorithms that are tailored to the unique characteristics of human and AI decisions in the current decision making task, thereby producing more accurate combined decisions than existing algorithms.

\subsection{On the Potential Misalignment between LLM Analysis and True AI Decision Rationales}
{%One limitation of this study is the potential misalignment between 
As discussed earlier, in our framework, the analysis produced by the LLM on a decision making task does not necessarily align with the actual decision rationale of the AI model (e.g., the random forest models used in this study). 
Since the LLM is not directly informed of the internal workings of the AI model (as we focus on scenarios where internals of AI models are not accessible in this study), its analysis---generated based on general knowledge about the task---may not capture the specific decision boundaries or feature correlation relationships of the AI model. 
However, we note that in this study, accurately explaining the AI model's decision rationale is not the primary motivation for including the LLM-powered analysis. 
Instead, LLM-powered analysis is used to provide a subjective interpretation of the AI model's recommendation and prompt decision makers to engage in critical thinking when there is no access to the actual explanations of the AI model. That is, while the AI model serves as the primary advisor for human decision makers and provides them with the decision recommendation, the LLM serves as the secondary advisor supplementing the primary advisor by providing its own justifications to the primary advisor's recommendation, which allows the decision maker to put the recommendation into context.

We argue that the potential lack of alignment between the LLM-powered analysis and the AI model's true decision rationale may not be a concern in many cases. First, the main motivation for including the LLM-powered analysis in our framework is to encourage decision makers' critical reflection of the decision task as well as the AI recommendation. Even if these analyses deviate from the AI model's true decision rationale, it could still effectively draw decision makers' attention to key features related to the decision, thereby guiding decision makers' independent and more thoughtful evaluation of the recommendation, allowing them to act on it cognitively rather than blindly trusting/not trusting it simply due to the lack of transparency. 
Second, in many scenarios, there may exist multiple reasoning paths to arrive at the same recommendation, making it less practical to align the LLM analysis with the ``true'' decision rationale of the AI model, which may not even be well-defined. In fact, even when actual AI explanations can be obtained, established explainable AI methods were often found to have limited fidelity~\cite{miro2024assessing}, and different methods can provide different explanations for the same decision of the same model~\cite{krishna2022disagreement,lai2019many}. Thus, when the actual AI explanations are not accessible, the LLM-powered analysis could just be viewed as one possible reasoning path to arrive at the AI model's recommendation when having the LLM engage in ``perspective taking'' to rationalize that recommendation, or it could even be viewed as the LLM's independent (and true) reasoning path when it has to arrive at the AI model's recommendation. 
The degree to which the LLM's reasoning path looks reasonable may provide critical insights into the validity of the AI recommendation, as the perceived reasonableness of the LLM's reasoning may correlate with the plausibility and robustness of the AI recommendation. 
%produced by having the LLM engage in ``perspective taking'' to rationalize the AI model's recommendation could at least be viewed as one of the possible reasoning paths to arrive at that recommendation.  
Finally, when the ultimate goal is to improve the decision maker's appropriate reliance on the AI model and thus increase their decision accuracy, the exact reasoning behind the recommendation of the AI model might not matter as long as the information provided by the secondary advisor (i.e., the LLM) leads to a better-informed decision. %In our algorithmic framework, this is achieved by strategically selecting the LLM analysis to be presented to nudge the decision maker towards making a target decision that is likely more accurate than the human or AI's independent decision. 
In this sense, compared to the precise content of the LLM-powered analysis, the knowledge about to what extent presenting a LLM analysis will nudge decision makers towards making a desirable target decision is more critical for %the proposed algorithmic framework to 
effectively improving humans' decision accuracy. In our algorithmic framework, this knowledge is captured through our human behavior model.

That said, we acknowledge that the when helping decision makers gain accurate understandings of the internal workings of the AI model is a primary end-goal, the LLM-powered analyses may bring about risks as humans may build an inaccurate mental model of the AI's internal workings based on these analyses. In extreme cases, the LLM-powered analyses may even ``sugercoat'' incorrect AI recommendations or hide ethical issues underneath the AI model, such as model biases~\cite{slack2020fooling}. To address this risk, the ultimate solution is to increase the transparency of the AI model to obtain the actual explanations of the model, and the proposed algorithmic framework could still be used to determine how to present these explanations selectively and progressively. However, without access to actual AI explanations, methods should be designed to increase people's awareness of the potential mismatch between the LLM analysis and the true AI decision rationale. Moreover, one may consult multiple secondary advisors (e.g., multiple LLMs) to analyze the AI recommendation and triangulate the reasoning process; this may help the human decision makers understand the diversity of possible interpretations of the AI recommendation and reduce the likelihood of being misled by the misinterpretation of any single secondary advisor.

\subsection{Design Implications for Human-LLM Interaction}

Our study demonstrates that while LLMs can generate and provide informative analysis for human decision makers, how to present this information is critical to its effective utilization. The heuristic design of interactions between humans and LLMs, when not carefully curated, often proves inefficient and fails to achieve the intended positive utility of LLM's analytical capabilities. For example, when decision makers are directly supplied with an abundance of LLM-generated information,  the information overload would overwhelm users and potentially result in decision fatigue, making it difficult for users to identify relevant information quickly. In addition, the practice of randomly slicing abundant information into pieces or relying solely on standard importance metrics to guide the presentation of data does not adequately consider the cognitive processes of how humans process such information. Such methods may lead to prolonged interactions between humans and LLMs, %continuous streams of LLM-generated information 
which may also overwhelm and confuse users, leading to suboptimal engagement and diminished utility of the LLM outputs. 

To mitigate these issues and enhance the practical utility of LLM for users, it is essential to integrate considerations of cognitive and contextual factors into the design of interaction paradigms between humans and LLMs to facilitate more effective and efficient interaction. For example, one important consideration in designing these interaction paradigms is to determine the most valuable information to present to users from the large pool of content that LLMs can generate. Given that LLMs are adept at producing vast quantities of information, ranging from seemingly meaningful to less relevant content, it is crucial to implement intelligent selection strategies to group information based on decision making priorities or estimated human cognitive needs. This may allow the LLM to dynamically adjust to the user’s immediate needs and contexts by  predicting what information is most pertinent based on user behavior and feedback. Additionally, allowing users to customize the presentation and management of information within the interface can be another promising approach to explore in the future. Customization options might include adjusting the volume, complexity, and format of the information to better align with individual processing styles and needs.  Finally, incorporating continuous feedback loops within the interface design is crucial for optimizing the interaction between humans and LLMs. These feedback loops enable users to provide input on the usefulness of the information presented, which can inform and refine the algorithms that select and present data, ensuring that the LLM remains dynamically aligned with user needs and preferences.

\subsection{Generalization of Methods and Findings}\label{sec:generalization}
%Despite the insightful results of enhancing appropriate reliance in AI-assisted decision making by our proposed framework, we acknowledge that our findings may not directly apply to other types of decision making tasks and subject populations. 

We acknowledge that our study has a few important limitations regarding the generalizability of our methods and findings. First, our study focused on two specific decision making tasks: income prediction and recidivism prediction. These tasks are widely used in previous research on AI-assisted decision making~\cite{van2021effect,wang2021explanations,green2019disparate,bansal2019updates,zhang2020effect} and feature a tabular data format with an explicit structure of features; the property of these tasks allows us to effectively apply LLM in the analysis and estimate how humans might react to these analyses. The success of our proposed framework in these two different tabular-data-based tasks strengthens our confidence in its potential to generalize to other AI-assisted decision making scenarios involving tabular data. 
However, applying our framework to decision making tasks with different data types, such as vision-based or text-based decision making tasks, presents additional challenges and requires further adaptation. This is because the image or text data do not contain explicit structured information that is amenable to analysis by the LLM in the same manner as tabular data.
%requires further adaptation. Specifically, tasks involving data in image or text modalities that do not feature the normalized explicit structure amenable to analysis by the LLM in the same manner as tabular data. This presents a challenge for direct application of our proposed framework.  
One potential solution is to convert these unstructured data types into a structured format that fits our proposed framework. For example, in text-based tasks, one could first use an LLM to extract semantically meaningful information from the text (e.g., text sentiment, key subjects in the text). %analyze the text from different pre-specified aspects and generate corresponding outputs. 
This extracted information
can then be treated as features, similar to how features are handled in tabular tasks, and subsequently input into the LLM for generating analyses. Likewise, for vision-based tasks, one could start by segmenting images into superpixels (i.e., groups of pixels representing visually meaningful entities)~\cite{achanta2012slic} or identifying relevant concepts in the images~\cite{kim2018interpretability}. %Each superpixel, which represents different information from various areas of the image, could then be analyzed by the LLM and yield the analyses. 
The presence and absence of certain superpixels and concepts would then serve as the features of the image, enabling LLMs to directly analyze them. 
The LLM analysis obtained could then be integrated into our framework, allowing it to work with a wider range of decision making tasks. When the transformation of unstructured data into structured formats is required before conducting the LLM-powered analysis, the decision on what features to be extracted from the data can be either made automatically (e.g., by the LLM) or manually by the human decision makers. Thus, how to ensure a comprehensive set of features will be extracted from the unstructured data becomes a critical challenge to be addressed.

% However, the applicability of our findings to other decision making domains, such as vision-based or text-based tasks, may be limited. Often, these domains do not feature the normalized, explicit structure amenable to analysis by LLMs in the same manner as tabular tasks. For instance, vision-based tasks frequently involve interpreting and analyzing the diverse textures and elements within images or videos, which requires different capabilities, such as image recognition and processing, that are distinct from the tabular data analysis typically handled by LLMs. Similarly, text-based tasks involve natural language processing, where the data structure is inherently less rigid and more contextual than tabular data. 

Secondly, as previously discussed, successfully nudging humans towards relying on AI models more appropriately hinges on the accurate modeling and prediction of how humans will react to different LLM-powered analyses. However, data on human-AI interaction collected in the past to train such behavioral models may not always align perfectly with current human behavior patterns, leading to potential discrepancies in effectiveness when these models are applied.   It is thus essential to continually update the human-AI interaction data. This update process ensures that the models can make predictions that align more accurately with current human behavior patterns.

In addition, in our current framework, we model %general patterns of human decision making process 
human decision makers' reactions to LLM-powered analysis on a population level 
without accounting for the unique characteristics of each individual. In other words, our human behavior model characterizes the behavior of an ``average'' decision maker. 
In our study, we found that the effects of the proposed algorithmic approach for selecting LLM-powered analysis in improving participants' final decision accuracy and enhancing their appropriate reliance on AI are robust across subpopulations with diverse demographic backgrounds and varying levels of AI knowledge, suggesting that modeling an average decision-maker is a reasonable modeling choice. 
That said, we acknowledge that this average modeling approach may neglect crucial individual differences that significantly affect the dynamics of human-AI interaction in AI-assisted decision making, and may indicate missing opportunities to further improve different individuals' decision performance by accounting for their unique characteristics. Future work could integrate various human characteristics (e.g., a person's intuition or prior knowledge about the task~\cite{chen2022machine}, need for cognition~\cite{buccinca2021trust}) into the human behavior models (i.e., one or more of the three model components---the initial state mapping model, the hidden state updating model, and the final decision model) to further accommodate individual preferences and traits. For instance, a person's competence or confidence in a specific decision task could be a critical moderating factor influencing how they would react to the AI recommendation and LLM-powered analyses. As individuals tend to exhibit low reliance on AI when they are more confident~\cite{mahmood2024designing}, explicitly accounting for human confidence in the behavior models may enable more efficient presentation of LLM-powered analyses (e.g., on tasks where humans are highly confident and the target decision suggests AI is not trustworthy, one may need to present fewer LLM analyses to nudge humans towards the target decision).  
%. Typically, individuals rely more on their own judgment in areas where they feel more confident, which could influence the nudge effectiveness of LLM analysis. 
As another example, a person's inherent tendency to trust AI or LLM systems can also be explicitly accounted for in the human behavior models, which may allow the algorithm to dynamically adjust which and how many LLM-powered analyses to be presented to the human decision makers based on their trust inclination. 
%For example, people with greater trust in an LLM may be more receptive to its analysis, thereby affecting the impact of nudging on humans to make targeted decisions.

Finally, we note that our study was conducted on Prolific, which primarily involved non-expert users in low-stake decision making scenarios. While this setting provided a suitable testbed for the evaluation of the appropriateness of human trust in AI-assisted decision making, we urge caution should be used when generalizing our conclusions to other populations or decision making scenarios. For example, in high-stake decision making scenarios where decision makers may utilize different cognitive strategies and where the consequences of errors are more significant, it is unclear whether the intelligent interaction paradigms we designed for interactions between humans and LLMs will perform equally well. %under these conditions. 
However, we believe that if sufficient human-AI interaction data can be collected in high-stake scenarios to train highly accurate human decision making models, the potential to successfully nudge human decisions even in these critical environments still persists.

\subsection{Other Limitations}
Our study has a few additional limitations. For example, our LLM-powered analysis mainly relies on the GPT-4 model. Consequently, our results may not generalize to other fine-tuned LLMs that are specifically designed for decision making support in various specialized fields, such as medical LLMs employed in clinical decision making scenarios. The distinct capabilities and pre-designed functionalities of these specialized models could lead to different outcomes in human-LLM collaboration compared to those observed with GPT-4, which has general-purpose capabilities. Furthermore, in our study, we employed random forest models as the AI assistant to provide decision support. The outcomes observed could vary significantly with the use of different AI models with its own set of processing abilities, training datasets, and optimization goals, all of which could potentially influence the effectiveness and reliability of the decision making support provided.

\section{Conclusion}
In this paper, we present an initial exploration of whether and how incorporating LLM-powered analysis can enhance the performance of human-AI teams in AI-assisted decision making, when explanations of the AI recommendations are not easily accessible or available. Through a randomized experiment, we first show that presenting LLM-powered analysis of each feature in decision making tasks, either sequentially or concurrently, does not significantly improve humans' performance in AI-assisted decision making. We then propose an algorithmic framework to characterize the effects of LLM-powered analysis on human decisions and dynamically decide which analysis to present. Our evaluation with human subjects shows that, by following the proposed approach, humans can achieve higher decision accuracy and exhibit reduced overreliance on AI in AI-assisted decision making. Overall, our study provides important experimental evidence regarding the effectiveness of incorporating LLMs in AI-assisted decision making, and how to design intelligent interaction methods between humans and LLMs to fully unlock the potential of LLMs for promoting better human-AI collaboration in decision making.

\section*{Acknowledgments}
We thank the support of the National Science Foundation under grant IIS-2229876 and IIS-2340209 on this work. Any opinions, findings, conclusions, or recommendations expressed here are those of the authors alone.

% In this paper, we present a study to obtain a systematic understanding of whether and how much value people attach to AI assistance, and how the incorporation of AI assistance in writing workflows changes people's writing perceptions and performance. Our results show that people are willing to forego financial payments to receive writing assistance especially if AI can provide direct content generation assistance and the writing task is highly creative. In addition, our results highlight that although the generative-AI powered assistance is found to offer benefits in increasing people's productivity and confidence in writing, direct content generation assistance offered by AI also comes with risks,
% including decreasing people’s sense of accountability and diversity in writing. Our work provides important implications for the design of human-AI collaborative writing systems, and we hope our findings in this paper can spur more explorations in the domain of human-AI co-writing.

\bibliographystyle{ACM-Reference-Format}
\bibliography{reference}

%%
% %% If your work has an appendix, this is the place to put it.
% \clearpage
% \appendix

% \label{sec:appendix}

% \counterwithin{figure}{section}
% \counterwithin{table}{section}
% \input{appendix}

\end{document}